\documentclass[aps,pre,twocolumn,superscriptaddress,showpacs] {revtex4}
\usepackage[latin1]{inputenc} 
\usepackage{amssymb}
\usepackage{epsfig}
\usepackage{amsmath}
\usepackage{times}
\usepackage[colorlinks,linkcolor=blue,anchorcolor=blue,citecolor=blue]{hyperref}
\usepackage{color}
\usepackage{graphicx}
\begin{document}

\title{A rigorous and efficient approach to finding and quantifying symmetries in complex networks}

\author{Yong-Shang Long}
\affiliation{State Key Laboratory of Precision Spectroscopy and School of Physics and Electronic Science, East China Normal University, Shanghai 200241, China}

\author{Zheng-Meng Zhai}
\affiliation{State Key Laboratory of Precision Spectroscopy and School of Physics and Electronic Science, East China Normal University, Shanghai 200241, China}

\author{Ming Tang} \email{tangminghan007@gmail.com}
\affiliation{State Key Laboratory of Precision Spectroscopy and School of Physics and Electronic Science, East China Normal University, Shanghai 200241, China}
\affiliation{Shanghai Key Laboratory of Multidimensional Information Processing, East China Normal University, Shanghai 200241, China}

\author{Ying Liu}
\affiliation{School of Computer Science, Southwest Petroleum University, Chengdu 610500, P. R.China}

\author{Ying-Cheng Lai}
\affiliation{School of Electrical, Computer and Energy Engineering, Arizona State University, Tempe, AZ 85287, USA}

\begin{abstract}

Symmetries are fundamental to dynamical processes in complex networks such as cluster synchronization, which have attracted a great deal of current research. Finding symmetric nodes in large complex networks, however, has relied on automorphism groups in algebraic group theory, which are solvable in quasipolynomial time. We articulate a conceptually appealing and computationally extremely efficient approach to finding and characterizing all symmetric nodes by introducing a structural position vector (SPV) for each and every node in the network. We prove mathematically that nodes with the identical SPV are symmetrical to each other. Utilizing six representative complex networks from the real world, we demonstrate that all symmetric nodes can be found in linear time, and the SPVs can not only characterize the similarity of nodes but also quantify the nodal influences in spreading dynamics on the network. Our SPV-based framework, in additional to being rigorously justified, provides a physically intuitive way to uncover, understand and exploit symmetric structures in complex networks.

\end{abstract}
\date{\today}
\maketitle

\section{Introduction}

Symmetries are ubiquitous in natural systems. In physics, the existence of
a symmetry implies the conservation of a physical quantity and, as such, a
great deal can be learned about the system without the need of analyzing
the system details. The same principle has been applied to complex networks for
identifying and understanding intricate dynamical phenomena such as cluster
synchronization~\cite{SIG:2003,PLM:2014,SFP1:2016,SFP2:2016,DRF:2020},
which otherwise would be difficult to analyze. Given a complex network,
a symmetry implies the existence of a group or a cluster of structurally
completely equivalent nodes. When certain dynamical process occurs on the
network, the nodes in a symmetric cluster, due to their complete equivalence,
tend to be more readily synchronized among themselves than with nodes outside
the cluster. A purely random network, such as the Erd\"{o}s-R\'{e}nyi
network~\cite{ER:1959}, typically has zero symmetry in the sense that the
probability is zero for any two nodes to be equivalent in terms of their
connection structure in the network. However, networks in the real world are
not purely random but, in fact, they often possess a large number of
symmetries~\cite{MBD:2009,MBS:2008,XYX:2008,PLM:2014} that affect not only
properties of the network such as spectrum, redundancy and robustness, but
also the various dynamical processes on the network. For example, in
the brain network, symmetries play an important role in the network dynamics
such as remote synchronization and singularity~\cite{NVV:2013,KLT:2019}. For
network computations, symmetries can be exploited to coarse-grain and reduce
the dimensionality of the network, merge symmetric nodes, and generate the
so-called entropy graph of the original network~\cite{SFS:2019} with reduced
computational complexity~\cite{XYM:2008,SR:2020}. In complex networks, the
mechanisms by which symmetries arise include replicative growth such as
duplication~\cite{CFL:2003}, evolution from basic principles~\cite{XYX:2008},
and functional optimization~\cite{NTM:2016}.

In order to take advantage of the network symmetries, an essential task is to
find all the symmetrical nodes. Existing methods are based on the algebraic
group theory, such as dividing the equivalence class by backtracking search
based on the coloring theory~\cite{MBD:1976}. A commonly used algorithm is
NAUTY - a set of procedures for efficiently determining the automorphism group
of a vertex-colored graph~\cite{MBD:1978}. A closely related problem is the
classical graph isomorphism problem in theoretical computer science - to
determine if two graphs are structurally identical or isomorphic. It was
proved recently by Babai~\cite{Babai:2016} that graph isomorphism is solvable
in quasipolynomial time: $N^{P(\log{N})}$ for a graph of $N$ nodes, where
$P(\cdot)$ is some polynomial. This means that, theoretically the graph
isomorphism is almost efficiently solvable. While the existing algorithms
are capable of finding the symmetric nodes, to quantify the symmetries
associated with different nodes remains to be an open problem. In particular,
in network science, a variety of centrality measures have been introduced to
classify and characterize the importance of nodes, such as the degree,
eigenvector centrality~\cite{BP:1987}, betweenness~\cite{FLC:1977},
H-index~\cite{HJE:2005}, PageRank~\cite{Newman:book}, k-core~\cite{DSN:2006},
etc., based on which concepts such as spectral coarse-grained networks have
been developed~\cite{GDD:2007,GDD:2008}. For a set of symmetric nodes, the
values of these centrality measures are identical. In terms of the network
symmetries, an important piece of knowledge that has been missing is to
uncover a statistical measure to quantify the degree of symmetries among
different nodes. The discovery of such a measure may also lead to more
efficient algorithms for finding the symmetric nodes than the existing
automorphism group based methods.

In the paper, we introduce an efficient statistical measure to characterize
the network symmetries. By defining a structural position vector (SPV) for
each and every node in the network, we exploit algebraic group theory to prove
rigorously that having the same SPV for a set of nodes is sufficient to
guarantee their symmetry. We demonstrate, using six large complex networks
from the real world, that {\em all} the known symmetrical structures can be
found accurately. In fact, since the computations required involve only a
small number of multiplications between the adjacency matrix (that is
typically sparse) and a vector, the symmetries can be found in linear time.
We then propose and demonstrate an SPV-based index to describe
the structural similarity among the nodes, which provides an effective
algorithm for coarse-graining the network. We further articulate a centrality
measure to quantify the nodal spreading influence - the role played by the
nodes in the epidemic spreading dynamics, which is validated using real-world
networks. Our SPV-based framework not only introduces a computationally
extremely efficient method to find all the symmetric structures but also
leads to a meaningful way to quantify the nodal symmetries, providing
deeper insights into the symmetries of complex networks.

\section{Results}

We first define SPVs and describe their relationship with nodal symmetries.
We then use the finite-order SPVs to find the symmetric structures of six
real-world networks and demonstrate how the SPVs can be used to describe the
structural similarity of nodes. Finally, we propose an SPV-based centrality
to quantify the nodal propagation influence.

\subsection{SPVs and nodal symmetries}
For nodes that are completely symmetrical to each other in the network, their
structural positions are identical. First, we set an initial or the $0$th-order
structural position $l_i^0 = 1$ for each node in the network, where $i$ is the
nodal index. For all nodes in the network, their $0$th-order structural
positions can be represented by a vector: $\mathbf{L}^0=[1,1,\cdots,1]^T$.
That is, initially, we disregard the edges and assume that all nodes have
the same structural positions. To take into account the edges, we note that
the structural position of a node is related to the structural positions of
its $n$th order neighbors, which is determined by the $n$th power of the
network adjacency matrix $\mathcal{A}^n$. Multiplying $\mathbf{L}^0$ from the
left by $\mathcal{A}^n$, we get $\mathbf{L}^n = \mathcal{A}^n\cdot\mathbf{L}^0$,
where $\mathbf{L}^n$ ($n=1,\ldots,\infty$) is a vector containing information
about the $n$th order structural positions of all nodes in the network, whose
$i$th component is given by $L^n_i=\sum_{j=1}^{N}[A^n]_{ij}L_j^0$, which is
the sum of the number of paths of length $n$ to node $i$ from all nodes in
the network. Specifically, the $1$st order structural position of a node is
its degree and the $\infty$th order is nothing but the eigenvector centrality
of the node. Note that the vectors $\mathbf{L}^n$ ($n=1,\ldots,\infty$) contain
information about the structural positions of all nodes, but it is not the
SPV that needs to be node-specific. To define the SPV for node $i$, we use
the $i$th component in $\mathbf{L}^n$, for $n=1,\ldots,\infty$:
$\mathbf{L}_i \equiv (L^1_i,L^2_i,\cdots,L^{\infty}_i)$. While the vector
$\mathbf{L}_i$ is infinite-dimensional, for a finite network of size $N$,
making the structural position vector $\mathbf{L}_i$ $N$-dimensional suffices.
We thus have $\mathbf{L}_i \equiv (L^1_i,L^2_i,\cdots,L^{N}_i)$.

Making use of the algebraic group theory, we can prove rigorously that if the
SPVs of nodes $i$ and $j$ are identical, i.e.,
$(L^1_i,L^2_i,\cdots,L^{N}_i)=(L^1_j,L^ 2_j,\cdots,L^{N}_j)$, then the two
nodes are symmetrical to each other in the network. Details of the proof are
presented in {\bf Methods}.

\begin{figure} [ht!]
\includegraphics[width=\linewidth]{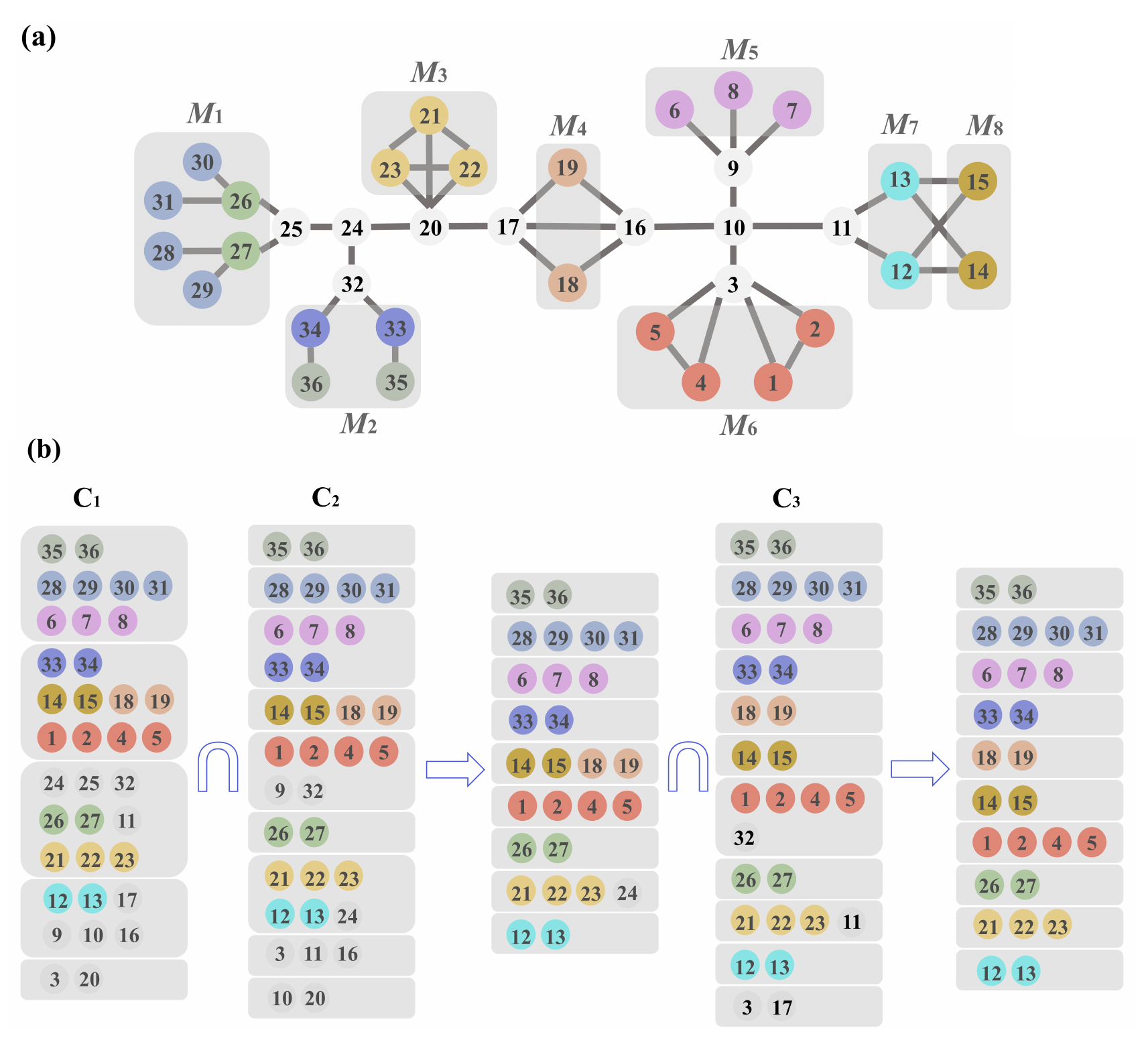}
\caption{Schematic representation of finding the symmetrical nodes using the
SPVs.  (a) A toy network containing eight symmetric motifs. (b) Nodal
classification $\mathbf{C}_n$ via the single $n$th-order vector $\mathbf{L}^n$,
for $n = 1,2,3$, for each node in (a). The intersections of the nodal
classification yield the symmetric nodes.}
\label{fig:toy}
\end{figure}

Figure~\ref{fig:toy} presents a simple example to illustrate how the symmetric
nodes can be found using the SPVs. The toy network has eight symmetric motifs
$M_i$ ($i=1,2\cdots,8$), as shown in Fig.~\ref{fig:toy}(a). In motif $M_1$,
there are two sets of nodes that are symmetric to each other: $\{28,29,30,31\}$
and $\{26,27\}$. A set of symmetric nodes form an orbit, e.g., nodes 26 and 27.
Motif $M_1$ thus consists of two orbits with two and four symmetric nodes,
respectively, motif $M_3$ has one orbit of three symmetric nodes, and so on.
We say that $M_1$ and $M_3$ are represented by the orbits
$O_2\stackrel{2}{\longrightarrow}O_4$ and $O_3^*$, respectively
[{\bf Supplementary Note (SN) 1}]. The orbital representations of the eight
symmetric motifs in Fig.~\ref{fig:toy}(a) and their corresponding geometric
factors are listed in Table~{\bf S1}. The leftmost column in
Fig.~\ref{fig:toy}(b) is a diagram of five blocks determined according to the
vector $\mathbf{L}^1$,
where all the nodes in a block have identical component values
in $\mathbf{L}^1$.
Similarly, the columns under $\mathbf{C}_2$ and $\mathbf{C}_3$ have nine
and 11 blocks, respectively. The middle column represents the set of
intersection between the sets under $\mathbf{C}_1$ and
$\mathbf{C}_2$,
which is done according to the criterion that all nodes
in a block must have identical component values in both $\mathbf{L}^1$ and
$\mathbf{L}^2$, although the common value in $\mathbf{L}^1$ may not equal
that in $\mathbf{L}^2$.
The rightmost column in Fig.~\ref{fig:toy}(b) consists of distinct blocks
obtained through the interaction among the sets corresponding to
$\mathbf{C}_1$, $\mathbf{C}_2$, and $\mathbf{C}_3$, which already contain all
the symmetric nodes of the network as specified in Fig.~\ref{fig:toy}(a).
This means that, for this toy network of 36 nodes, three
iterations of the initial vector $\mathbf{L}^0$ already suffice to yield all
the symmetric nodes.

We apply the principle to finding the symmetrical motifs in real-world
networks. We consider six such networks: two information networks
Odlis.net~\cite{RJM:2014} and p2p-Gnutella~\cite{RMF:2002}, the social network
Facebook~\cite{MJJ:2012}, a cooperative network CA-GrQc~\cite{LJK:2007},
a biological network Yeast~\cite{BDZ:2003}, and a voting network
Wiki-Vote~\cite{LJH:2010}, whose structural properties are listed in
Table~\ref{tab:real_world_networks}. Through the SPVs of the nodes, we succeed
in finding all the symmetrical motifs in six representative real-world
in linear time.
The results are shown in Table~{\bf S2} and Fig.~{\bf S2}.

\begin{table*} [ht!]
\begin{center}
\caption{Structural properties of six real-world networks.}
\vspace*{0.0in}
\label{tab:real_world_networks}
\begin{tabular}{|p{.15\textwidth}|p{.10\textwidth}|p{.10\textwidth}|p{.10\textwidth}|p{.10\textwidth}|p{.10\textwidth}|p{.10\textwidth}|}
\hline
	Networks   &   $|V|$  &   $|E|$  &  $ \langle k \rangle$  &  $ \langle d \rangle$  & $c$   & $r$\\ \hline
	Facebook   &   4039   &   88234  &        43.69           &      3.69              & 0.606 & 0.064 \\ \hline
	Odlis.net  &   2909   &   16388  &        11.27           &      3.17              & 0.296 & -0.173 \\ \hline
	CA-GrQc    &   4158   &   13422  &        6.46            &      6.05              & 0.557 & 0.639 \\ \hline
	p2p-Gnutella   &   6301   &  20779 &       6.60           &     4.64              & 0.011 & 0.035 \\ \hline
	Yeast      &   2224   &   6609  &          5.94           &      4.38              & 0.138 & -0.105 \\ \hline
	Wiki-Vote  &   7066   &   100736  &        28.51           &     3.25              & 0.142 & 0.083 \\ \hline	
\end{tabular}
\end{center}
\end{table*}

\subsection{Identification of symmetric nodes using SPVs}
Theoretically, when calculating the SPVs of a network of size $N$, it is
necessary to obtain all $\mathbf{L}^n$ vectors for $n=1,\ldots,N$. However,
if the network has a small diameter (as for many real-world
networks~\cite{XSW:2017}), $\mathbf{L}^n$ tends to converge fast, typically
requiring only a few iterations of $(\mathcal{A}^n\cdot \mathbf{L}^0)$.
This means that, computationally, our method for finding the symmetric nodes
can be extremely efficient. To quantify the computational efficiency, we
define the following accuracy measure $\rho$ to find the smallest $n$ value,
denoted as $n^*$, for which all symmetric structures in the network can
be completely identified through the first $n$th-order SPVs:
\begin{equation} \label{eq:rho}
\mathbf{\rho} =\frac{min\{U,Q\}}{max\{U,Q\}}\frac{1}{U}\sum_{k=1}^{U}\frac{n_k}{N_k},
\end{equation}
where $U$ is the number of non-trivial orbits in the symmetric structure of
the network (see {\bf SN 1} for a definition of ``orbits'') and
$Q$ is the
number of nontrivial classes obtained through
$\mathbf{C}_1\cap \mathbf{C}_2 \cap \ldots \cap \mathbf{C}_n$
[e.g., the distinct blocks in the rightmost column in Fig.~\ref{fig:toy}(b)],
where ``nontrivial'' means that there are at least 2 elements in the class.
The quantity $n_k$ ($k=1,\ldots,U$) is the number of nodes in the $k$th orbit
of the network, and $N_k$ is the number of nodes in the class (block)
containing the $k$th orbit. By definition, we have $\rho \leq 1$. The larger
the value of $\rho$ is, the higher is the accuracy of the identified symmetric
nodes. The perfect case $\rho = 1$ means that all symmetric nodes can be
identified using the first $n$ orders SPVs.

\begin{figure} [ht!]
\includegraphics[width=\linewidth]{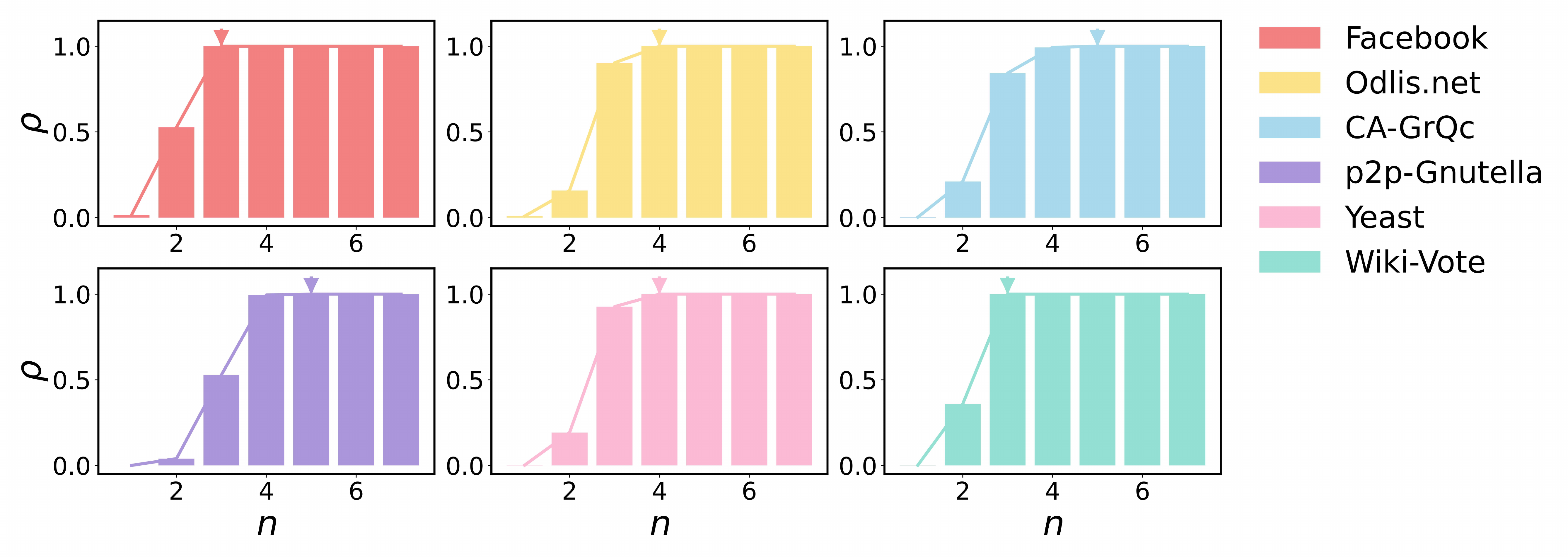}
\caption{ Accuracy measure $\rho$ of identifying all symmetric nodes versus
the order-$n$ SPVs for the six real-world networks in
Tab.~\ref{tab:real_world_networks}. For Facebook and Wiki-Vote, the values
of $n^*$, the minimum order required to have all symmetric nodes in the
network correctly identified, are three. For Odlis.net and Yeast network,
$n^*=4$. For CA-GrQc and p2p-Gnutella networks, $n^* = 5$. The very small
values of $n^*$ across all the networks is strong indication that the SPV-based
framework is computationally extremely efficient for finding all the symmetric
nodes in large complex networks.}
\label{fig:rho}
\end{figure}

Figure~\ref{fig:rho} shows the accuracy measure $\rho$ versus $n$ for the six
real-world networks listed in Tab.~\ref{tab:real_world_networks}. It can be
seen that $100\%$ accuracy as characterized by $\rho = 1$ is achieved for
$n\geq 5$ for all six networks. For Facebook and Wiki-Vote network, this
is achieved even for $n\geq 3$. The value of $n^*$ is thus exceedingly small
for all cases tested, indicating that all the symmetric nodes can be found
extremely efficiently -in time that is only proportional to the size of the
network (in linear time).

\subsection{Quantifying nodal structure similarity based on SPVs}
The similarity measure between two nodes based on their common neighbors has
been widely used in problems such as link prediction and recommendation
systems~\cite{LLJ:2009}. In general, two nodes with a similar
connection structure often have similar values of multiple centrality measures
such as degree, H-index, k-core, eigenvector centrality, and betweenness
centrality,  which is also important to coarse-graining networks and node
influence identification. The use of a single centrality measure is usually
not sufficient to describe the structural similarity between different nodes,
as two nodes can have identical values for certain centrality but differ
dramatically in other measures. (For example, nodes 10 and 12 in
Fig.~\ref{fig:toy}(a) have the same degree but their betweenness centrality
values are quite different.) Symmetrical nodes are expected to have exactly
the same SPV and centrality measures~\cite{XYM:2008}. Intuitively, two nodes
with similar (not equal) SPVs are structurally similar, so we define the
following similarity index $r$, the Euclidean distance between the structural
position vectors of nodes $i$ and $j$:
\begin{equation} \label{eq:r}
	r = \sqrt{\sum_{n=1}^N(\tilde{L}_i^n-\tilde{L}^n_j)^2},
\end{equation}
where $\tilde{L}_i^n$ ($n=1,2,\cdots,N$) is the $n$th component of the SPV
of node $i$, normalized by the largest $n$th component value among all the
nodes: $\tilde{L}_i^n = L_i^n/\max (L^n_1,L^n_2,\cdots,L_N^n)$. A small value
of $r$ indicates a relatively high degree of similarity between the structures
of nodes $i$ and $j$. The results in Fig.~\ref{fig:rho} indicate that SPVs of
size at most about five are needed to accurately identify all symmetric nodes
in the six real-world networks. It thus suffices to use the following truncated
SPV with six components:
$\tilde{\mathbf{L}}^*_i \equiv (\tilde{L}_i^1,\tilde{L}^2_i,\cdots,\tilde{L}_i^6)^T$,
to calculate the similarity measure $r$.

\begin{figure*} [ht!]
\includegraphics[width=\linewidth]{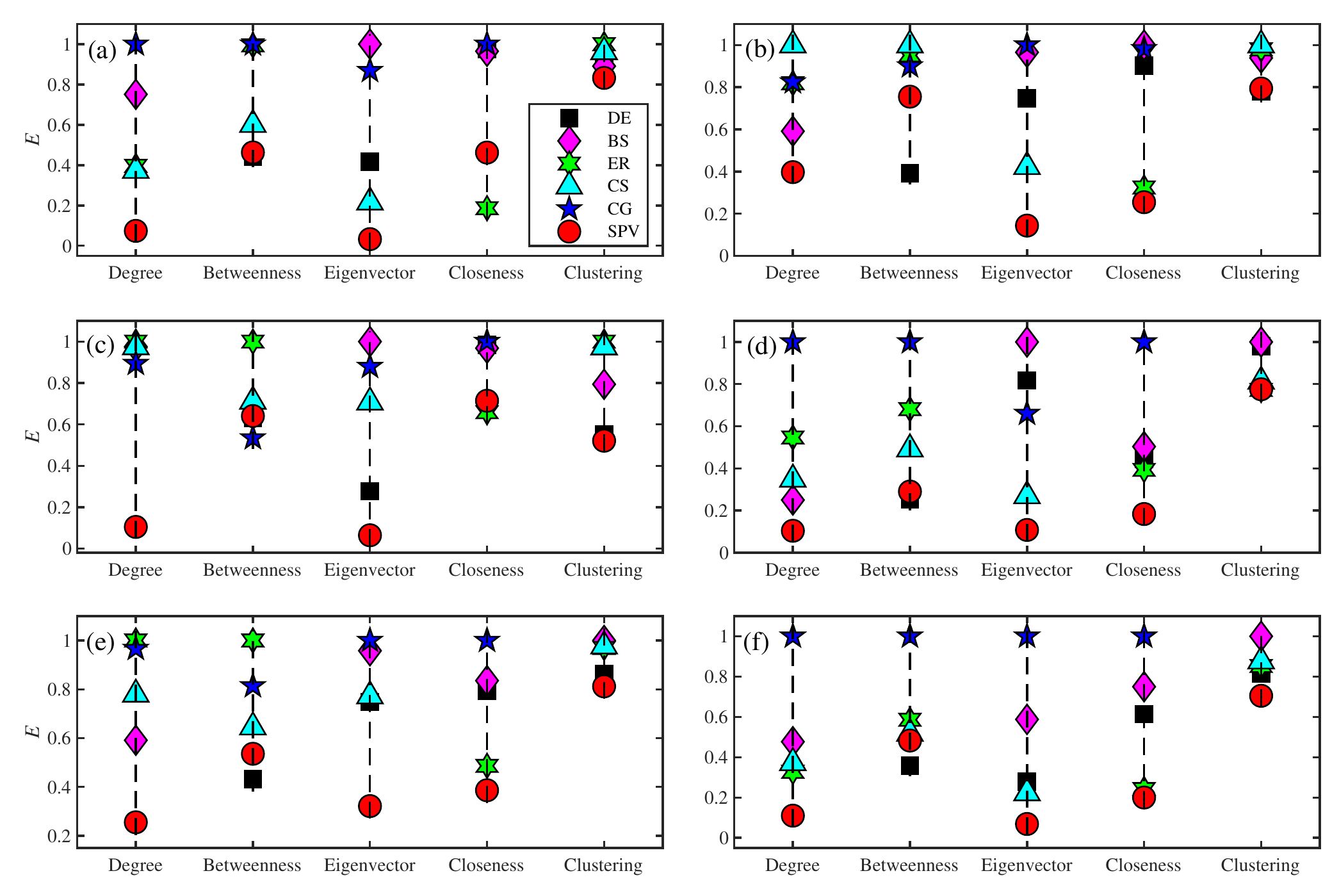}
\caption{Relative errors of node clustering by different clustering methods.
(a-f): Respective results from six real-world networks: Facebook, Odlis.net,
CA-GrQc, p2p-Gnutella, Yeast and Wiki-Vote. Black squares, pink diamonds, light
green hexagon stars, cyan triangles, blue pentagrams and red circles represent
the relative errors associated with the abscissa when clustering nodes based
on degree (DE), betweenness (BS), eigenvector (ER), closeness centrality (CS),
clustering centrality (CG) and structural position vector (SPV), respectively.}
\label{fig:error}
\end{figure*}

As an application, we consider the problem of clustering nodes with similar
structures. Our point is that the method based on the similarity index $r$
defined in Eq.~(\ref{eq:r}) performs better than many previously known methods.
To demonstrate this, we use the $k$-mean algorithm to cluster the nodes in
a complex network into $\tilde{N}$ classes. To verify the effect of clustering, we coarse-grain the network and use the properties of the center of each
class to approximate the properties of nodes in this class. For a comparative
analysis, we consider five clustering methods based on degree, betweenness,
eigenvector centrality, closeness centrality, and clustering coefficient,
respectively. We define the coarse-grained error value $E$ as
\begin{equation}
E = \sum_{c=1}^{\tilde N} \sum_{j=1}^{n_c} \sqrt{(z_{c_j}-Z_c)^2},
\end{equation}
where $\tilde N$ is the number of clusters, $n_c$ is the number of nodes in
the $c$th cluster, $z_{c_j}$ is a conventional statistical measure of
node $j$ in the $c$th cluster, and $Z_c$ is the statistical measure of the
center of the $c$th cluster. Given a network, we cluster all the nodes using
each of the six clustering methods. Each clustering method produces errors for
the five conventional statistical measures. Except for the clustering method
based on similarity index $r$, only the errors of the four statistical measures
are calculated in the other five clustering methods. For example, if we
coarse-grain the network using the nodal degree, we calculate the errors from
betweenness, eigenvector centrality, closeness centrality, and clustering
coefficient. (In this case, the error from the nodal degree must be minimum
by definition, so including this error in the comparison is not meaningful.)
We then normalize these coarse-grained error values to obtain the relative
errors. Figure~\ref{fig:error} shows, for the six real-world networks in
Tab.~\ref{tab:real_world_networks}, the relative errors for the coarse-grained
networks obtained from each of the six clustering methods, where the
coarse-grained scale is $\tilde{N} = N(k)$, with $N(k)$ being the scale of
coarse-grained network based on nodal degree. It can be seen that the relative
node-clustering errors associated with three or four statistical measures
are minimized when clustering nodes based on similarity index $r$.
Especially the relative errors from degree, eigenvector centrality,
and clustering coefficient are the smallest for all networks. In addition,
the relative error of closeness centrality is the smallest for Odlis.net,
p2p-Gnutella, Yeast, and Wiki-Vote when the coarse-grained network based on
similarity index $r$. While in the remaining two networks, the relative error
from closeness centrality can reach the second smallest although it is not
the smallest. No conventional clustering method can generate node clustering
as accurate as our similarity measure $r$, indicating that the SPVs can better
describe the structural similarity of nodes than the conventional methods.

\subsection{Quantifying role of nodes in propagation dynamics based on SPV}
Propagation is a fundamental type of dynamical processes
in real-world networks, with examples ranging from epidemic spreading in
human social networks~\cite{APS:2021,FMI:2019} and diffusion of crisis in
financial networks~\cite{STC:2017,LZH:2020} to cascading failures in power
grids~\cite{YYN:2017,SBW:2018} and signal transmission in neural
networks~\cite{AKA:2018,RAG:2018}.
Measuring/quantifying the influence of nodes in propagation and identifying
the nodes that play a critical role in the processes are issues with
significant basic and applied values in network science. Conventionally, nodal
centralities such as the degree, H-index and k-core are used for these
purposes. The principle of coarse graining stipulates that nodes with the
same degree, H-index or k-core have identical influences. Our result is that
the SPVs provide an alternative and potentially more powerful way to measure
the nodal influences in the propagation dynamics. This is based on the
intuition that the more similar the structures of nodes, the closer are their
influences.

Following our approach to quantifying nodal structure similarity, We cluster
the network using the similarity index $r$ and calculate the two-norm value
$b_i$ of the truncated SPVs $\tilde{\mathbf{L}}_i^*$ of the center of each
class. This leads to the cluster centrality $b_i$ based on the SPVs, which
can be used to quantify the influence of nodes in this class. When the network
is not clustered, $b_i$ is the SPV centrality.

\begin{figure*} [ht!]
\includegraphics[width=\linewidth]{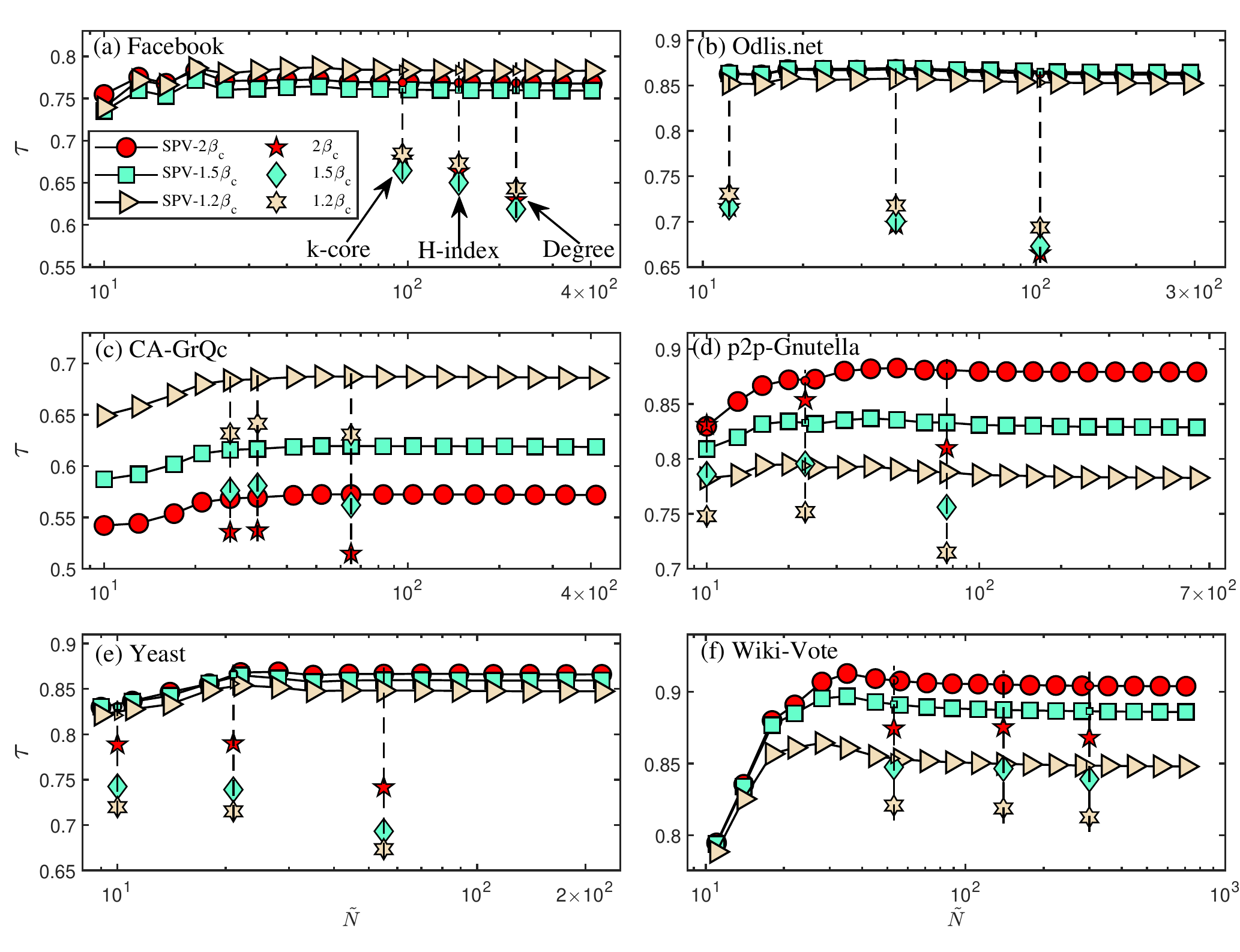}
\caption{Kendall's $\tau$ correlation coefficient for the six real-world
networks. Shown is the $\tau$ value between the SPV-based cluster centrality
and the nodal influence $R$ versus $\tilde{N}$, the number of clusters in the
SPV-based coarse-grained network for $0.001N \leq \tilde N \leq 0.1N$, with
$N$ being the size of the whole network. The light yellow triangles, light
green squares, and red circles, respectively, represent the correlation
coefficient $\tau$ between the SPV-based cluster centrality and the node
influence $R$ for infection rate $\beta =1.2\beta_c$, $1.5\beta_c$, and
$2\beta_c$ in the SIR dynamics. For comparison, the $\tau$ values between three
conventional centralities(degree, H-index, and k-core) and $R$ are shown. (For
each conventional centrality, the corresponding coarse-grained network has a
single value of $\tilde{N}$.) The light yellow six-pointed star, the light
green diamond, and the red five-pointed star, respectively, represent the
$\tau$ value between the three conventional centralities and $R$ for
$\beta =1.2\beta_c$, $1.5\beta_c$, and $2\beta_c$.}
\label{fig:tau}
\end{figure*}

To demonstrate the SPV-based quantification of nodal influences in
propagation dynamics, we use the classic SIR
(susceptible-infected-refractory) model. In the simulations, we set each
node $i$ is as the origin of the spreading dynamics and calculate the
fraction $R_i$ of the recovered nodes. We average over 5000 independent runs
to obtain the mean $R_i$ value that characterizes the propagation influence
of node $i$. We thus have two sequences for all nodes in the network: the
cluster centrality sequence $(b_1,b_2,\cdots,b_N)$ and the propagation
influence sequence $(R_1,R_2,\cdots,R_N)$, so their correlation can be
calculated by using, e.g., the Kendall's correlation coefficient $\tau$,
where $-1\leq \tau \leq 1$ (see {\bf SN 2} for a detailed definition). A large
value of $\tau$ indicates a higher accuracy of the centrality in ranking the
nodal influences. Fig.~\ref{fig:tau} shows, for the six real-world networks,
Kendall's correlation coefficient between different cluster centralities and
the propagation influence as $\tilde N$, the number of clusters increases. It
can be seen that the cluster centralities are generally suitable for ranking
the nodal influences. For example, for the network Odlis.net,
Fig.~\ref{fig:tau}(b) gives $\tau \geq 0.85$ and, for the Yeast network in
Fig.~\ref{fig:tau}(e), we have $\tau \geq 0.8$. Within a certain range, as the
number of clusters increases, the correlation increases correspondingly. When
the number of clusters is sufficiently large (e.g., $\tilde N \ge 50$), the
correlation value plateaus. For comparison, we also calculate the correlation
coefficient between the propagation influence sequence and each of the three
conventional coarse-graining methods: degree, H-index, and k-core, where each
method gives only a single value for each network. As shown in
Fig.~\ref{fig:tau}, for all six networks, the ranking performance of the
SPV-based cluster centrality is consistently and significantly better than
those of the three conventional methods in characterizing the nodal influences
in epidemic spreading.

As the number of clusters exceeds about 50 ($\tilde N \geq 50$), the cluster
centrality achieves a high accuracy in ranking the nodal influence. Note that,
for $\tilde N = N$, the cluster centrality becomes the actual SPV centrality.
Let $\mbox{SPV}(\tilde{N})$ denote the cluster centrality for the
coarse-grained network of $\tilde{N}$ clusters, where
$\mbox{SPV} \equiv \mbox{SPV}(\tilde N = N)$. To systematically compare the
nodal ranking performances of $\mbox{SPV}$ and $\mbox{SPV}(\tilde{N})$ with
those of the six conventional nodal ranking methods (i.e., those based on
the degree, H-index, k-core, closeness, betweenness, and eigenvector
centrality), we choose $\tilde N$ to be the number of clusters obtained from
the H-index. The results for the six real-world networks are listed in
Tab.~\ref{tab:tau}, where the basic parameter $\beta$ (the infection rate)
associated with SIR dynamics is set to be $\beta = 1.2\beta_c$ with $\beta_c$
being the propagation threshold value ({\bf SN 2}). We see that $\mbox{SPV}$
and $\mbox{SPV}(\tilde{N})$ have approximately the same accuracy that is
generally higher than the accuracies of the conventional nodal ranking methods.
There are a small number of exceptions. For example, in the p2p-Gnutella
network, the eigenvector and closeness centralities have higher correlation
with the propagation influence than that with the SPV-based methods. This is
due to the fact that the network has no apparent local structures. Additional
results for $\beta = 1.5\beta_c$ and $\beta = 2\beta_c$ can be found in
Tables~{\bf S3} and {\bf S4}, respectively. We find that, except for the two
networks p2p-Gnutella and CA-GrQc, the SPV-based cluster centralities have the
best ranking performance. [In the CA-GrQc network, there is a high clustering
coefficient $c$ and assortativity $r$, indicating an core-periphery
structure~\cite{KSM:2018}. As a result, indices that reflect the importance
of the core nodes (such as k-core, closeness centrality, eigenvector
centrality) can better predict the spreading influence of nodes.]

In general, the influence of a node depends on the adjacency of different
orders, corresponding to the components of each order in the SPV. The
contribution of each order of adjacency is different~\cite{LYT:2016}.
Motivated by this, we further study the relative importance of each order
component in the SPV to the influence of the node. Specifically, when
clustering the nodes, we adjust the dimension of the truncated SPV and the
weight of each order component. As shown in Figs.~{\bf S2} and {\bf S3},
within a certain range, the more dimensions that the truncated SPV possesses,
the better is the achieved ranking performance. Figures {\bf S4} and {\bf S5}
show that, in a network with a small average path length, the importance of
$\mathbf{L}^2$ is larger than that of $\mathbf{L}^3$. In networks with a
larger average path length, the conclusion is the opposite. (Details can be
found in {\bf SN 3}).

\begin{table*} [ht!]
\begin{center}
\caption{Kendall's $\tau$ correlation coefficient between node
influence $R$ and eight indices for $\beta=1.2\beta_c$.}
\vspace*{0.0in}
\label{tab:tau}
\begin{tabular}{|p{.13\textwidth}|p{.09\textwidth}|p{.09\textwidth}|p{.09\textwidth}|p{.09\textwidth}|p{.13\textwidth}|p{.12\textwidth}|p{.09\textwidth}|p{.09\textwidth}|}
\hline
	Networks   & Degree & H-index &  k-core  &  Closeness & Betweenness & Eigenvector & SPV & SPV($\tilde N$) \\ \hline
	Facebook   &   0.6434   & 0.6729  &  0.6846  &  0.4286    & 0.4331   & 0.6511 & \bf{0.7869}  & 0.7834    \\ \hline
	Odlis.net  &   0.6939   & 0.7178  &  0.7309  &  0.7801    & 0.5406   & 0.8307 & 0.8514  & \bf{0.8577}    \\ \hline
	CA-GrQc    &   0.6307   & 0.6413  &  0.6318  &  0.5806    & 0.3587   & 0.5965 & \bf{0.6856}  & 0.6847    \\ \hline
	p2p-Gnutella   & 0.7148   &  0.7516 & 0.7480 &  0.8309    & 0.6533   & \bf{0.8452} & 0.7825  & 0.7939    \\ \hline
	Yeast      &   0.6734   & 0.7151  &  0.7197  &  0.8216    & 0.5646   & 0.8272 & 0.8476  & \bf{0.8556}    \\ \hline
	Wiki-Vote  &   0.8126   & 0.8188  &  0.8207  &  0.8257    & 0.7431   & 0.8508 & \bf{0.8509}  & 0.8498    \\ \hline	
\end{tabular}
\end{center}
\end{table*}

\section{Discussion}

Symmetric structures are fundamental to dynamical processes on complex
networks, and accurately identifying these structures is of great importance.
While the existing, algebraic-group theory based methods can find all the
symmetric nodes in the network in quasipolynomial time, these methods do not
yield a quantification or differentiation of the degree of symmetries.
Two outstanding questions are: (1) does a method in linear time
exist for finding all the symmetric structures of large complex networks?
and (2) how to quantify the degree of nodal symmetries?
This paper addresses both questions through the development of a unified
framework. The fundamental concept that we articulate to accomplish this task
is the nodal structural position vectors (SPVs). In particular, we define the
SPV through the adjacency relationships in the network. Mathematically, by
employing the algebraic group theory, we transform the relationship of equal
SPVs into an eigenvector problem to rigorously prove that nodes with equal SPVs
must be symmetrical. Utilizing real-world networks, we demonstrate that our
SPV-based method allows all symmetric structures to be found with a small
number of iterations of the SPV in computation time that is proportional to
the network size. More importantly, our SPV-based framework enables the role
of nodal symmetries in dynamical processes to be quantified and differentiated.

The similarity between two nodes based on common neighbors has been widely
used in link prediction and recommendation systems. Since the symmetrical
nodes have exactly the same structural characteristics, our SPV framework
provides a natural way to define the structural similarity. In particular,
defining a structural similarity index or centrality, we cluster the nodes in
the network and find that the SPV-based clustering method is remarkably
effective in coarse-graining the network in that it outperforms the previous
clustering methods based on the traditional centralities such as the degree,
eigenvector centrality, H-index, closeness and betweenness centralities.
More importantly, from the standpoint of dynamical processes, our SPV-based
clustering centrality can be used to measure and quantify the roles of nodes
in propagation dynamics. Detailed calculation using six real-world networks
indicates that, in most cases, the SPV-based centrality outperforms the
conventional centralities by a large margin in predicting the nodal influences
on propagation dynamics.

Taken together, our work provides a conceptually appealing and computationally
extremely efficient framework to find symmetric nodes in large complex networks,
totally bypassing the sophisticated steps in the conventional methods based on
automorphism groups. The method also leads to a statistical characterization
of nodal symmetries, with direct applications to coarse-graining of complex
networks and cluster synchronization that may occur among remote nodes.

\section{Methods}

The main mathematical result of this paper is the theorem: {\em Nodes with
equal SPVs must be symmetrical to each other.} Here we prove this theorem by
using the algebraic group theory.

Mathematically, the existence of symmetric nodes in the network stipulates
an automorphism $\pi$ satisfying the property that $\{u,v\}$ is an edge
in the network if and only if $\{\pi(u),\pi(v)\}$ is also an edge. All the
automorphisms of a network constitute an automorphism group $G(\pi)$.

First, we prove two lemmas.

\textbf{Lemma 1} The permutation $\pi$ is an automorphism if and only if
$\mathcal{P}\cdot\mathcal{A}=\mathcal{A}\cdot\mathcal{P}$ ,where $\mathcal{A}$,
$\pi$ and $\mathcal{P}=(p_{ij})$ are the adjacency matrix, permutation, and
permutation matrix of the network, respectively. When nodes $i$ and $j$
have a permutation relation with each other: $v_i=\pi(v_j)$, the corresponding
elements in the permutation matrix are $p_{ij}=p_{ji}=1$, and all other
elements are zero.

\underline{Proof}: Assume  $v_h=\pi(v_i),v_k=\pi(v_j)$, then
\begin{equation}
\begin{split}
\left \{\begin{array}{ll} (\mathcal{P}\cdot\mathcal{A})_{hj}=\sum p_{hl}a_{lj}=a_{ij}, \\
(\mathcal{A}\cdot\mathcal{P})_{hj}=\sum a_{hl}p_{lj}=a_{hk}, \end{array} \right. \end{split}\end{equation}
i.e., $\mathcal{P}\cdot\mathcal{A}=\mathcal{A}\cdot\mathcal{P}$ if and only if
$a_{ij}=a_{hk}$. That is, the permutation $\pi$ is an automorphism. $\blacksquare$

\textbf{Lemma 2}  If there is a permutation matrix $\mathcal{P}$ such that
$\mathcal{P}\cdot\mathbf{x}=\pm \mathbf{x}$, where $\mathbf{x}$ is an
eigenvector of the adjacency matrix of the network, then the permutation
corresponding to $\mathcal{P}$ is an automorphism.

\underline{Proof}: Let $\mathcal{A}\cdot\mathbf{x}=\lambda \mathbf{x}$,
where $\lambda$ and $\mathbf{x}$ are the eigenvalue and the associated
eigenvector of the network adjacency matrix $\mathcal{A}$, respectively. Assume
$\mathcal{P}\cdot\mathbf{x}=\pm \mathbf{x}$. Since the permutation matrix
$\mathcal{P}$ is derived from the identity matrix through a series of
elementary transformations, $\mathcal{P}$ is invertible. Left multiplying
both sides of the characteristic equation
$\mathcal{A}\cdot\mathbf{x}=\lambda \mathbf{x}$ by $\mathcal{P}$, we get
$\mathcal{P}\cdot \mathcal{A}\cdot\mathbf{x}=\lambda\mathcal{P}\cdot\mathbf{x}$.
With $\mathcal{P}\cdot\mathbf{x}=\pm \mathbf{x}$, we get
\begin{equation}
\mathcal{P}\cdot \mathcal{A}\cdot\mathbf{x} = \pm \lambda \mathbf{x}
= \pm \mathcal{A}\cdot\mathbf{x} = \mathcal{A}\cdot \mathcal{P}\cdot\mathbf{x},
\end{equation}
which gives
$(\mathcal{P}\cdot\mathcal{A}-\mathcal{A}\cdot\mathcal{P})\cdot\mathbf{x} = 0$.
Because $\mathcal{A}$ is an $N\times N$ symmetric matrix, there must be $N$
linearly independent eigenvectors, denoted as $\mathbf{x}_1$, $\mathbf{x}_2$,
$\cdots$, $\mathbf{x}_N$. The matrix
$\mathcal{X}\equiv(\mathbf{x}_1,\mathbf{x}_2,\cdots,\mathbf{x}_N)$ has full
rank. We thus have
$(\mathcal{P}\cdot\mathcal{A}-\mathcal{A}\cdot\mathcal{P})\cdot\mathcal{X}=0$.
Expressing the matrix
$(\mathcal{P}\cdot\mathcal{A}-\mathcal{A}\cdot\mathcal{P})$
in a row form, we get
\begin{equation}
(\mathcal{P}\cdot\mathcal{A}-\mathcal{A}\cdot\mathcal{P})\cdot\mathcal{X} =
\left[\begin{matrix}\phi_1 \\ \phi_2 \\ \vdots \\ \phi_N \end{matrix} \right]\mathcal{X}=\left[ \begin{matrix} 0 \\0 \\ \vdots \\ 0 \end{matrix} \right].
\end{equation}
Since matrix $\mathcal{X}$ has full rank, the equations $\phi_i \mathcal{X}=0$
have trivial solutions only: $\phi_i = 0$. We thus have
$\mathcal{P}\cdot\mathcal{A}=\mathcal{A}\cdot\mathcal{P}$. According to
Lemma 1, the permutation corresponding to the permutation matrix $P$ is
an automorphism. $\blacksquare$

\underline{\bf Outline of proof of main theorem}:
Assume that the SPVs of the nodes $x$ and $y$ are equal:
$\mathbf{L}^n_x=\mathbf{L}^n_y$ for $n=1,2,\cdots,N)$. Because $\mathcal{A}$
is a symmetric matrix, it must have $N$ linearly independent eigenvectors. The
initial vector $\mathbf{L}^0=(1,1,\cdots,1)^T$ can then be linearly represented
by the eigenvectors $\bm{\eta}_i$ for $i=1,2,\cdots,N$:
$\mathbf{L}^0=\sum_{i=1}^{N} t_i\bm{\eta}_i$. From
$\mathbf{L}^n = \mathcal{A}^n \cdot \mathbf{L}^0$, we get
\begin{equation}
\left \{
\begin{array}{ll}
	\mathbf{L}^1=\mathcal{A}\cdot\mathbf{L}^0=\mathcal{A}(\sum_{i=1}^{N}t_i\bm{\eta}_i)=\sum_{i=1}^{N}t_i\lambda_i \bm{\eta}_i,\\
	\mathbf{L}^2=\mathcal{A}^2\cdot\mathbf{L}^0=\mathcal{A}^2(\sum_{i=1}^{N}t_i\bm{\eta}_i)=\sum_{i=1}^{N}t_i\lambda_i^2 \bm{\eta}_i,\\
     \vdots,\\
	\mathbf{L}^N=\mathcal{A}^N\cdot\mathbf{L}^0=\mathcal{A}^N(\sum_{i=1}^{N}t_i\bm{\eta}_i)=\sum_{i=1}^{N}t_i\lambda_i^N \bm{\eta}_i.
\end{array}
\right.
\end{equation}
The equality $\mathbf{L}^n_x=\mathbf{L}^n_y$ for $n=1,2,\cdots,N$ gives
\begin{equation}
\left \{
\begin{array}{ll}
	\sum_{i=1}^N t_i \lambda_i \eta_{i,x}=\sum_{i=1}^N t_i \lambda_i\eta_{i,y},\\
	\sum_{i=1}^N t_i \lambda_i^2\eta_{i,x}=\sum_{i=1}^N t_i \lambda_i^2\eta_{i,y},\\
      \vdots\\
	\sum_{i=1}^N t_i \lambda_i^N\eta_{i,x}=\sum_{i=1}^N t_i \lambda_i^N \eta_{i,y}.
\end{array}
\right.
\end{equation}
or
\begin{equation}
\left \{
\begin{array}{ll}
	\sum_{i=1}^N t_i \lambda_i(\eta_{i,x}-\eta_{i,y})=0,\\
	\sum_{i=1}^N t_i \lambda_i^2(\eta_{i,x}-\eta_{i,y})=0,\\
      \vdots,\\
	\sum_{i=1}^N t_i \lambda_i^N(\eta_{i,x}-\eta_{i,y})=0.
\end{array}
\right.
\end{equation}
The determinant of this set of linear equations in $(\eta_{i,x}-\eta_{i,y})$ is
\begin{widetext}
\begin{equation}
C=\left|\begin{matrix}t_1\lambda_1 & t_2\lambda_2 & \cdots & t_N \lambda_N \\ t_1\lambda_1^2 & t_2\lambda_2^2 & \cdots & t_N \lambda_N^2 \\ t_1\lambda_1^3 & t_2\lambda_2^3 & \dots & t_N\lambda_N^3 \\ \vdots & \vdots & \ddots & \vdots\\ t_1\lambda_1^N & t_2\lambda_2^N & \cdots & t_N\lambda_N^N\end{matrix} \right|=t_1t_2\cdots t_N \lambda_1 \lambda_2\cdots \lambda_N\left|\begin{matrix} 1& 1 & \cdots & 1\\ \lambda_1 & \lambda_2 & \cdots & \lambda_N\\ \lambda_1^2 & \lambda_2^2 & \cdots & \lambda_N^2\\ \vdots & \vdots & \ddots & \vdots\\ \lambda_1^{N-1}& \lambda_2^{N-1}& \cdots& \lambda_N^{N-1}\end{matrix}\right|.
\end{equation}
\end{widetext}
which is the Vandermonde determinant. We have
$C=t_1t_2\cdots t_N \lambda_1 \lambda_2\cdots \lambda_N \prod_{1\leq j<i\leq N}(\lambda_i-\lambda_j)$.
If all eigenvalues of $\mathcal{A}$ are simple, zero is not its eigenvalue
and $t_i \neq 0$ for $i=1,2,\cdots,N$, then we have $C\neq 0$. The only
solution of $(\eta_{i,x}-\eta_{i,y})$ is zero, so
$\eta_{i,x}=\eta_{i,y}$. The permutation of nodes $x$ and $y$ implies that all
eigenvalues of $\mathcal{A}$ must satisfy $\mathcal{P}\cdot\bm{\eta}=\bm{\eta}$.
By Lemma 2, the permutation corresponding to the permutation matrix $P$ is an
automorphism, so the nodes $x$ and $y$ are symmetrical to each other.
$\blacksquare$

A more general and detailed proof is presented in {\bf SN 4}.

\section*{Data Availability}

All relevant data are available from the authors upon request.

\section*{Code Availability}

All relevant computer codes are available from the authors upon request.

\section*{Acknowledgments}

This work was supported by the National Natural Science Foundation of China
under Grant Nos.~82161148012, 11975099, 11575041, and the Science and
Technology Commission of Shanghai Municipality under Grant No.~14DZ2260800.
The work at Arizona State University was supported by the Office of Naval
Research through Grant No.~N00014-21-1-2323.

\section*{Author Contributions}
Y.-S.L., M.T., and Y.-C.L. designed research; Y.-S.L. performed research;
Y.-S.L., Z.-M.Z., M.T., and Y.-C.L. contributed analytic tools;
Y.-S.L., Z.-M.Z., M.T., and Y.L. analyzed data;
Y.-S.L., M.T., Y.L., and Y.-C.L. wrote the paper.

\section*{Competing Interests}

The authors declare no competing interests.

\section*{Correspondence}

To whom correspondence should be addressed. E-mail: tangminghan007@gmail.com.

\section*{References}

\bibliographystyle{naturemag}
\bibliography{Quantifyingsymmetry}

\end{document}



\begin{center}
{\large Supplementary Information for}
\end{center}

\begin{center}
	{\large\bf A rigorous and efficient approach to finding and quantifying symmetries in complex networks}
\end{center}

\begin{center}
{Yong-Shang Long, Zheng-Meng Zhai, Ming Tang$^*$, Ying Liu,
and Ying-Cheng Lai}
\end{center}

\newpage

\section*{Supplementary Note 1: Symmetry and automorphism groups}

A network is a graph $G(V,E)$ with vertex set $V$ and edge set $E$, where
a pair of nodes are adjacent if there is an edge between them. An automorphism
is a permutation of the vertices of the network, which preserves the adjacency.
The combinations of a set of automorphisms form a group
$\mathcal{G}=\mbox{Aut}(G)$ that describes the symmetries of the
network~\cite{BB:2013}. A network with a nontrivial automorphism group is
symmetric.

Consider the permutations of a set of $N$ nodes $Y={v_1,v_2,\cdots,v_N}$. The
support of a permutation $p$ is the set of points that $p$ moves:
$\mbox{supp}(p)=\{v_i\in V(G)| p(v_i)\neq v_i\}$. Two sets of automorphisms,
$P$ and $Q$, are support-disjoint if every pair of automorphisms $p\in P $ and
$q\in Q$ has disjoint support. Similarly, the automorphism subgroups $G_p$
and $G_Q$ generated by $P$ and $Q$ are support-disjoint if $P$ and $Q$ are
support-disjoint. The independent action of automorphism subgroups provides
a way to factorize the automorphism groups of complex networks into
``irreducible building blocks''~\cite{MBS:2008}. In particular, let $S$ be a
set of generators of the automorphism groups $\mbox{Aut}(G)$. Suppose that we
partition $S$ into $n$ support-disjoint subsets: $S=S_1\cup S_2 \cdots S_n$,
such that each $S_i$ cannot itself be decomposed into smaller support-disjoint
subsets. $H_i$ is called the subgroup generated by $S_i$ and every $H_i$
commutes with all others. A direct product decomposition of the automorphism
groups $\mbox{Aut}(G)$ is
\begin{equation} \label{eq:SN1_2}
	\mbox{Aut}(G)=H_1 \times H_2 \times \cdots \times H_n.
\end{equation}

\begin{figure} [ht!]
\includegraphics[width=\linewidth]{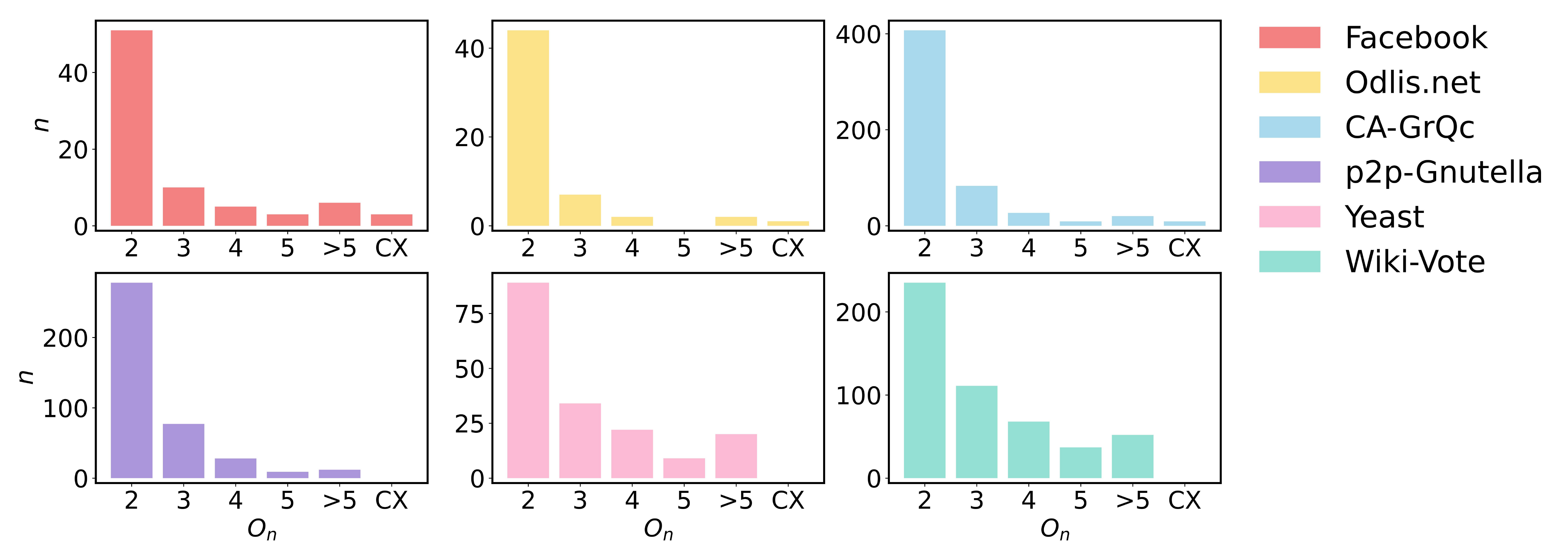}
\caption{All symmetric motifs in six real-world networks. $O_n$ is the number
of nodes contained in the symmetric motifs, and ``CX'' denotes the complex
symmetric motifs. The ordinate is the number of symmetric motifs.}
\label{fig:motif}
\end{figure}

\begin{table} [ht!]
\begin{center}
\caption{Orbital representation for the symmetric motifs and the corresponding
geometric factors. The meanings of the symbols are as follows: $O_\alpha^*$
specifies that the orbit contains $\alpha$ nodes that are connected with each
other; $O_\alpha $ means that the orbit contains $\alpha$ nodes and there is
no connection between them; $O_\alpha^\mu$ signifies that the orbit contains
$\alpha$ nodes with $\mu$ edges between them; and
$O_\alpha\stackrel{\gamma}{\longrightarrow}O_{\alpha'}$ means that the
symmetric motifs contain two orbits, and each node in the first orbit is
connected to $\gamma$ nodes in the second orbit. If $\gamma=\alpha'$, the
symmetric motif is decomposed into two independent symmetric motifs, e.g.,
$M_7$ and $M_8$ in Fig.~1(a). The group of all permutation of
$n$ objects is denoted as $S_n$, $C_n$ is the group of partial permutation
of $n$ objects, and $\wr$ is the wreath product~\cite{RJJ:2012}. Supplementary
Table~\ref{tabs2} lists that there are mainly two symmetric motifs in the six real-world
networks: $O_\alpha^* $ and $O_\alpha $, which are the basic symmetric motifs.
There are a few symmetrical motifs such as $O_\alpha^\mu$ and
$O_\alpha\stackrel{\gamma}{\longrightarrow}O_{\alpha'} $, which are the
complex symmetrical motifs.}
\vspace*{0.0in}
\label{tab:s1}
\begin{tabular}{|p{.2\textwidth}|p{.25\textwidth}|p{.2\textwidth}|}
\hline
	Symmetric Motif   &   Orbital Representation &   Geometric Factor  \\ \hline
	$M_1$  &   $O_2\stackrel{2}{\longrightarrow}O_4$   &  $S_2\wr S_2$   \\ \hline
	$M_2$  &   $O_2\stackrel{1}{\longrightarrow}O_2$   &   $S_2\wr S_1 \mbox{or}  S_2 $   \\ \hline
	$M_3$    &   $O_3^*$   &   $S_3$   \\ \hline
	$M_4,M_7,M_8$   &   $O_2$   &   $S_2$    \\ \hline
	$M_5$      &   $O_3$   &   $S_3$    \\ \hline
	$M_6$  &   $O_4^2$   &   $C_4$  \\ \hline	
\end{tabular}
\end{center}
\end{table}

The decomposition of automorphism group in Eq.~\eqref{eq:SN1_2} is unique and
irreducible~\cite{MBS:2008}. The direct product factorization given in
Eq.~\eqref{eq:SN1_2} is the geometric decomposition of $\mbox{Aut}(G)$ and each
factor $H_i$ a geometric component. The induced subgraph on a set of vertices
$X\subset V$ is the graph obtained by taking $X$ and any edges whose end points
are in $X$. The induced subgraph on the support of a geometric factor $H_i$ is
called the symmetric motif, denoted as $M_H$. A number of common symmetric
motifs in the real-world networks are shown in Fig.~1(a). The
orbital representation of the symmetric motifs therein and how the geometric
factors of the automorphism group are related to different symmetric motifs
are described in Supplementary Table~\ref{tab:s1}. For every vertex $v\in V$,
the set of vertices to which $v$ maps under the action of the automorphism
group $\mathcal G = \mbox{Aut}(G)$ is called a $G$-orbit of $v$, denoted as
$\Delta(v)$. More formally, we have
$\Delta (v)=\{g\cdot v\in V :g\in Aut(G)\}$. A symmetric motif generally
consists of single or multiple orbits.

\begin{table} [ht!]
\begin{center}
\caption{Symmetric motifs of the six real networks in the main text}
\vspace*{0.0in}
\label{tabs2}
\begin{tabular}{|p{.15\textwidth}|p{.775\textwidth}|}
\hline
	Network   &   Symmetric  Motif    \\ \hline
	Facebook  &   $(O_2^*)^{33}+(O_2)^{18}+(O_3^*)^9+O_3 +(O_4^*)^4+O_4+(O_5^*)^2+O_5+(O_7)^2+O_9+O_{11}+O_{13}+O_{14}+(O_4^2)^2+O_6^3$   \\ \hline
	Odlis.net  &   $(O_2^*)^{14}+(O_2)^{30}+(O_3)^7+(O_4)^2+O_8+O_{15}+(O_2\stackrel{1}{\longrightarrow} O_2^*)$  \\ \hline
	CA-GrQc     &   $(O_2^*)^{279}+(O_2)^{128}+(O_3^*)^{57}+(O_3)^{26}+(O_4^*)^{17}+(O_4)^{10}+(O_5^*)^7+(O_5)^2+(O_6^*)^4+(O_6)^3
                     +O_7^*+O_8^*+O_9^*+O_{10}^*+O_{12}^*+O_{13}^*+O_{14}^*+O_{15}^*+O_{16}^*+O_{17}^*+O_{18}^*+O_{23}^*+O_{32}^*+
                     (O_4^2)^7+O_6^6+O_{12}^{30}$  \\  \hline
	p2p-Gnutella  &   $O_2^*+(O_2)^{277}+(O_3)^{77}+(O_4)^{28}+(O_5)^{9}+O_6+(O_7)^3+O_8+O_9+(O_{10})^4+O_{26}+O_{32}$ \\ \hline
	Yeast      &     $(O_2^*)^2+(O_2)^{87}+(O_3)^{34}+(O_4)^{22}+(O_5)^{9}+(O_6)^9+(O_7)^5+(O_8)^3+O_9+(O_{10})^2$  \\ \hline
	Wiki-Vote  &     $(O_2)^{235}+(O_3)^{111}+(O_4)^{68}+(O_5)^{37}+(O_6)^{15}+(O_7)^{13}+(O_8)^{5}+(O_9)^{4}+(O_{10})^2
                     +(O_{12})^2+(O_{13})^2+O_{14}+O_{15}+O_{17}+O_{19}+O_{20}+O_{24}+O_{26}+O_{29}+O_{32}$ \\ \hline	
\end{tabular}
\end{center}
\end{table}

\section*{Supplementary Note 2: SIR model and Kendall's $\tau$ correlation coefficient}

\paragraph*{\bf SIR model.} The standard SIR (susceptible-infected-recovered)
model~\cite{PRC:2015} has been widely used in simulating epidemic spreading
and information propagation/diffusion dynamics. At each time step, an infected
node makes contact with its neighbors and each susceptible neighbor is
infected with the probability $\beta$. The infected node enters into the
state of recovery or removal with the probability $\lambda$. It is convenient
to set $\lambda=1$.

To quantify the spreading influence of node $i$, we initiate the spreading
process with $i$ being the infected seed and all other nodes being susceptible.
The spreading process stops when there are no infected nodes. We record the
fraction of recovered nodes $R_i$ as the spreading influence of the seed node.
According to the heterogeneous mean-field theory~\cite{NM6:2002,CCP:2010},
the epidemic threshold in the SIR model is given by
$\beta_c =\langle k\rangle/(\langle k^2\rangle -\langle k\rangle)$,
where $\langle k\rangle$ is the average degree and $\langle k^2\rangle$ is
the second moment of the random degree variable. In our simulations, we set
$\beta = 1.2\beta_c,1.5\beta_c,2\beta_c$. To eliminate the fluctuations of
$R_i$, we average the results over 5000 independent runs.

\paragraph*{\bf Kendall's $\tau$ correlation coefficient.} The Kendall's
$\tau$ correlation coefficient is a widely used statistical measure for
ranking the correlation. Consider two sequences associated with $|V|$ samples:
$x=(x_i,x_2,\cdots,x_{|V|})$ and $ y=(y_1,y_2,\cdots,y_{|V|})$. A pair of
orders $(x_i,x_j)$ and $(y_i,y_j)$ are concordant if both $x_i>x_j$ and
$y_i>y_j$ or if both $x_i<x_j$ and $y_i<y_j$, are discordant if $x_i>x_j$ and
$y_i<y_j$ or if both $x_i<x_j$ and $y_i>y_j$. If $x_i=x_j$ or $y_i=y_j$, the
pair is neither concordant nor discordant. As there are $|V|(|V|-1)/2$ pairs
of samples in total, the Kendall's $\tau$ correlation coefficient is defined as
\begin{equation} \label{eq:SN2_1}
\tau = 2(\delta^+ - \delta^-)/[|V|(|V|-1)],
\end{equation}
where $\delta^+$ and $\delta^-$ are the numbers of concordant and discordant
pairs, respectively.

\begin{table} [ht!]
\begin{center}
\caption{Kendall's $\tau$ correlation coefficient between node influence $R$ and eight indices for $\beta=1.5\beta_c$}
\vspace*{0.0in}
\label{tabs3}
\begin{tabular}{|p{.11\textwidth}|p{.07\textwidth}|p{.09\textwidth}|p{.07\textwidth}|p{.07\textwidth}|p{.09\textwidth}|p{.07\textwidth}|p{.07\textwidth}|p{.09\textwidth}|}
\hline
	Networks   & Degree & H-index &  k-core  &  Close-ness & Between-ness & Eigen-vector & SPV & SPV($\tilde N$) \\ \hline
	Facebook   &   0.6187   & 0.6501  &  0.6645  &  0.4633    & 0.4219   & 0.6715 & \bf{0.7633}  & 0.7602    \\ \hline
	Odlis.net  &   0.6729   & 0.7004  &  0.7160  &  0.7915    & 0.5203   & 0.8505 & 0.8640  & \bf{0.8693}    \\ \hline
	CA-GrQc    &   0.5620   & 0.5810  &  0.5762  &  0.6284    & 0.3273   & \bf{0.6477} & 0.6184  & 0.6168    \\ \hline
	p2p-Gnutella   & 0.7561   &  0.7957 & 0.7862 &  \bf{0.8813}    & 0.6910   & 0.8228 & 0.8286  & 0.8330    \\ \hline
	Yeast      &   0.6932   & 0.7390  &  0.7424  &  0.8345    & 0.5762   & 0.8121 & 0.8599  & \bf{0.8657}    \\ \hline
	Wiki-Vote  &   0.8392   & 0.8465  &  0.8476  &  0.8543    & 0.7624   & 0.8853 & \bf{0.8888}  & 0.8874    \\ \hline	
\end{tabular}
\end{center}
\end{table}

\begin{table} [ht!]
\begin{center}
\caption{Kendall's $\tau$ correlation coefficient between the nodal influence $R$ and eight indices for $\beta=2\beta_c$}
\vspace*{0.0in}
\label{tabs4}
\begin{tabular}{|p{.11\textwidth}|p{.07\textwidth}|p{.09\textwidth}|p{.07\textwidth}|p{.07\textwidth}|p{.09\textwidth}|p{.07\textwidth}|p{.07\textwidth}|p{.09\textwidth}|}
\hline
        Networks   & Degree & H-index &  k-core  &  Close-ness & Between-ness & Eigen-vector & SPV & SPV($\tilde N$) \\ \hline
	Facebook   &   0.6293   & 0.6635  &  0.6786  &  0.4955    & 0.4262   & 0.6579 & \bf{0.7712}  & 0.7685    \\ \hline
	Odlis.net  &   0.6651   & 0.6966  &  0.7160  &  0.7864    & 0.5101   & 0.8626 & 0.8632  & \bf{0.8678}    \\ \hline
	CA-GrQc    &   0.5146   & 0.5372  &  0.5358  &  \bf{0.6919}    & 0.3172   & 0.6624 & 0.5718  & 0.5695    \\
\hline
	p2p-Gnutella   & 0.8099   &  0.8537 & 0.8309 &  \bf{0.9183}    & 0.7385   & 0.7608 & 0.8794  & 0.8712    \\ \hline
	Yeast      &   0.7416   & 0.7900  &  0.7888  &  0.8152    & 0.6022   & 0.7689 & 0.8666  & \bf{0.8669}    \\ \hline
	Wiki-Vote  &   0.8680   & 0.8753  &  0.8744  &  0.8568    & 0.7789   & 0.8900 & 0.9038  & \bf{0.9051}    \\ \hline	
\end{tabular}
\end{center}
\end{table}

\section{Supplementary Note 3: Impact of dimension of truncated SPV and its component weights on nodal propagation influence}

In the main text, we find that clustering nodes based on the first six orders
of the truncated SPV $\tilde{\mathbf{L}}_i^*$ and the cluster centrality based on the
SPV can well predict the influence of nodes on propagation dynamics. The
influence of a node depends on its adjacency relationship at each order,
corresponding to the order components in the truncated SPV. The contributions
of the adjacency from different orders are different. Here we further study
the relative importance of each order component in the SPV to the nodal
propagation influence.  We first set the dimension of $\tilde{\mathbf{L}}_i^*$ to
be $n$: $\tilde{\mathbf{L}}^*_i = ({L}_i^1,{L}_i^2,\cdots,{L}_i^n)$ for
$n = 1,2,\cdots,10$. The scale of the coarse-grained network is set to
be $\tilde N = N(k)$, $N(\mbox{H-index})$, and $N(\mbox{k-core})$,
respectively. To be representative, we consider two cases where the
propagation rate is above the threshold: $\beta = 1.2\beta_c$ and
$1.5\beta_c $. We then assign weights to different order components of
$\tilde{\mathbf{L}}_i^*$: $\tilde{\mathbf{L}}^*_i=(w_1{L}_i^1,w_2{L}_i^2,\cdots,w_n{L}_i^n)$.

\begin{figure} [ht!]
\centering
\includegraphics[width=0.8\linewidth]{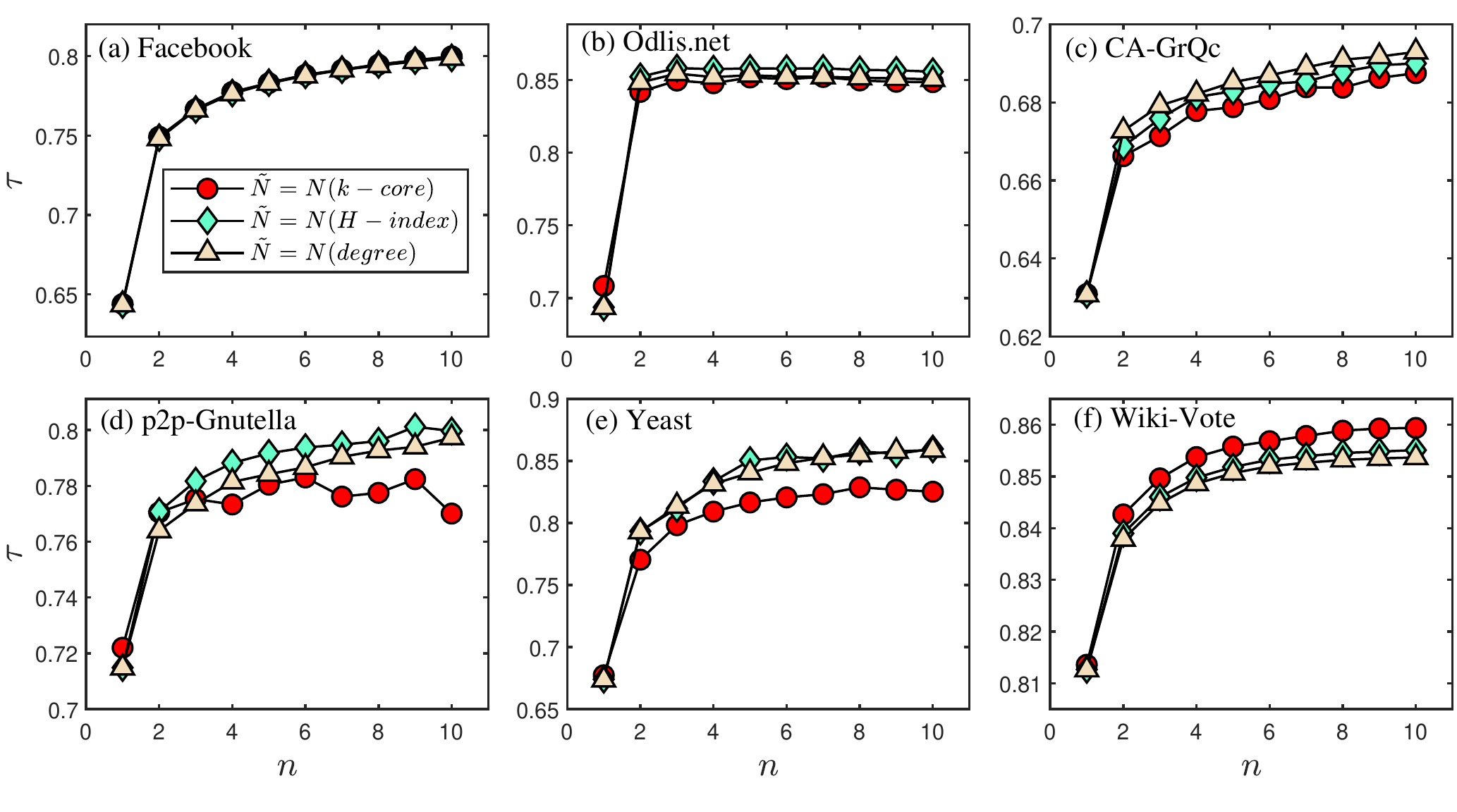}
\caption{Kendall' $\tau$ correlation coefficient between the SPV-based cluster
centrality and nodal influence $R$ as a function of the dimension $n$ of the
SPV for $\beta = 1.2\beta_c$.}
\label{fig:Dim1}
\end{figure}

\begin{figure} [ht!]
\centering
\includegraphics[width=0.8\linewidth]{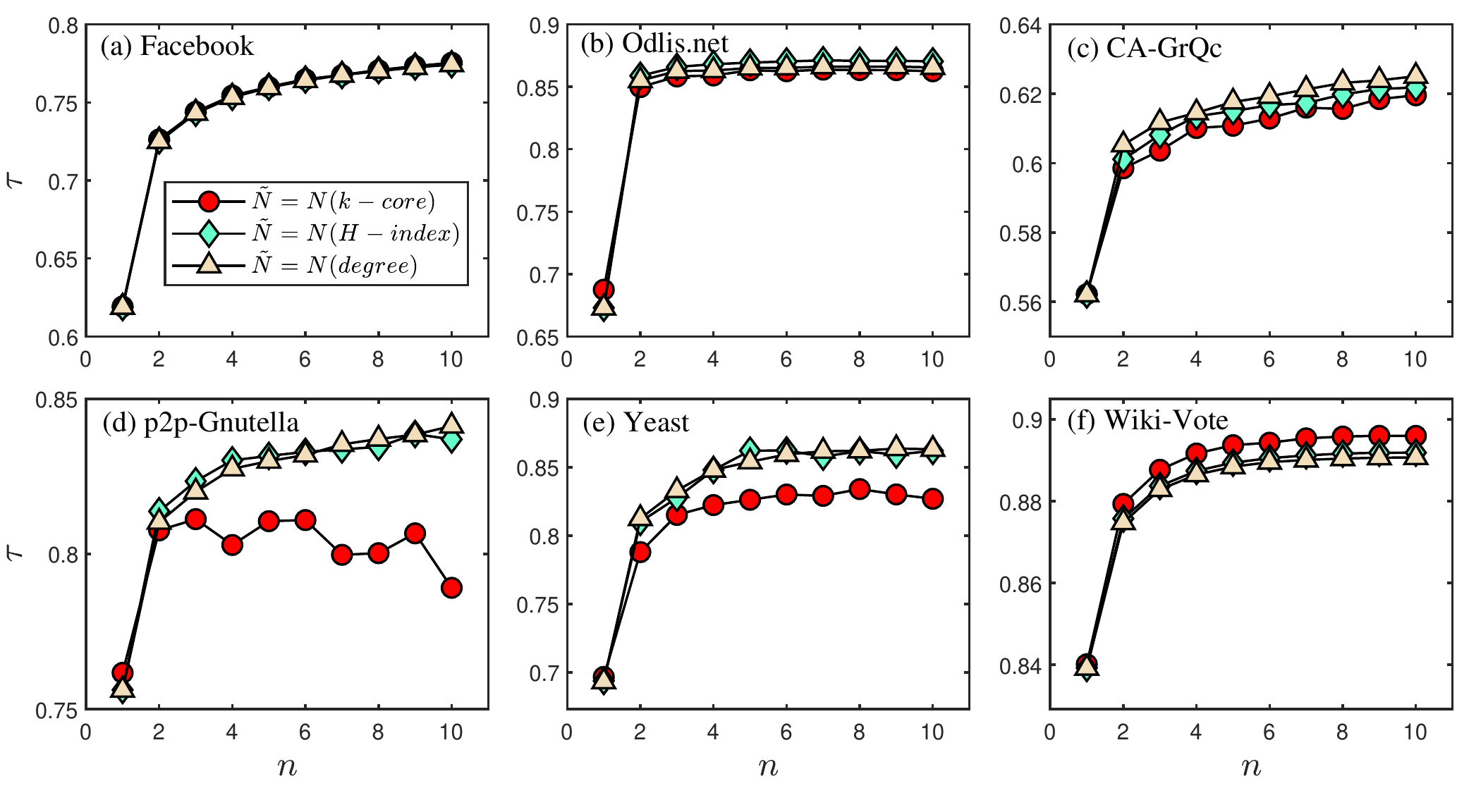}
\caption{Kendall' $\tau$ correlation coefficient between the SPV-based cluster
centrality and nodal influence $R$ as a function of the dimension $n$ of the
SPV for $\beta = 1.5\beta_c$.}
\label{fig:Dim2}
\end{figure}

Figures~\ref{fig:Dim1} and \ref{fig:Dim2} show that, when using the SPV-based
cluster centrality index to measure the influence of a node, the performance
improves with the dimension of the SPV and then saturates. For example, if the
SPV is one-dimensional, the only information it contains is the nodal degree,
so the performance is relatively poor. However, a three-dimensional SPV can
already predict the nodal influence well, but a higher dimensional SPV only
leads to incremental improvement in the performance. In Figs.~\ref{fig:Dim1}(d)
and \ref{fig:Dim2}(d), for $\tilde N=N(k-core)$, the performance curve is
irregular, due mainly to the reason that the scale of the coarse-grained
network is too small so the performance of the SPV based cluster centrality is
not stable. In this case, increasing the dimension of the SPV does not improve
the performance significantly. When the scale of the coarse-grained network is
large, as the dimensionality increases, the level of classification becomes
more detailed. In this case, the performance of cluster centrality tends to
improve with the dimension before reaching a plateau.

\begin{figure} [ht!]
\centering
\includegraphics[width=0.8\linewidth]{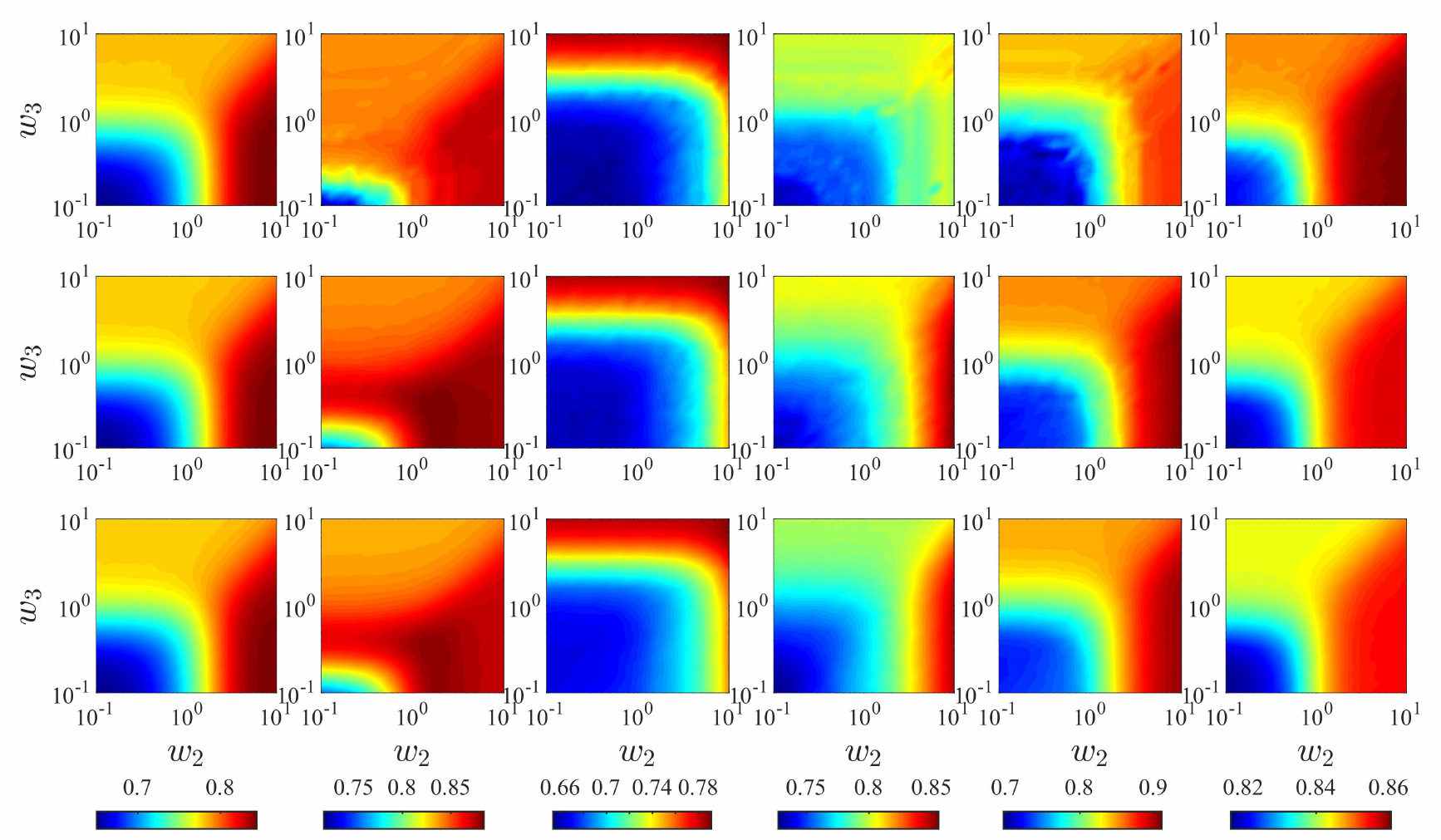}
\caption{Effect of SPV component weights on nodal propagation influence. Shown
is Kendall's $\tau$ correlation coefficient between the SPV-based cluster
centrality and nodal influence $R$ versus the weights of the second and third
components of the truncated SPV (i.e., the second- and third-order structural
positions), denoted as $w_2$ and $w_3$, respectively, for $\beta = 1.2\beta_c$. From left to right, the networks in each column are Facebook, Odlis.net,
CA-GrQc, p2p-Gnutella, Yeast, and Wiki-Vote. The scale of the coarse-grained
network in each row from top to bottom is $\tilde N = N(\mbox{k-core})$,
$N(\mbox{H-index})$ and $N(k))$.}
\label{fig:Wei1}
\end{figure}

\begin{figure} [ht!]
\centering
\includegraphics[width=0.8\linewidth]{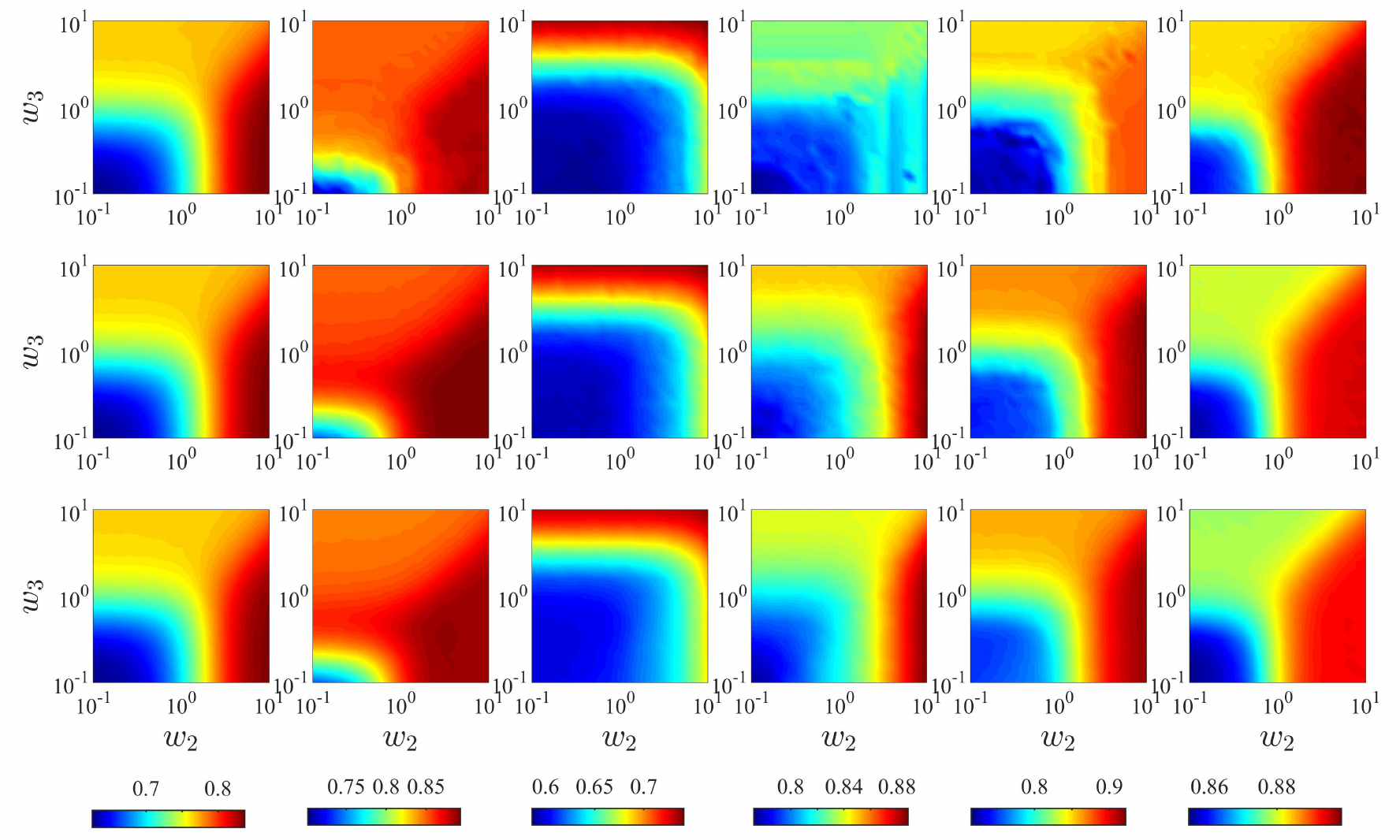}
\caption{Effect of SPV component weights on nodal propagation influence.
The legends are the same as those in Fig.~\ref{fig:Wei1} except that the
value of parameter $\beta$ is different: $\beta=1.5\beta_c$.}
\label{fig:Wei2}
\end{figure}

We also study the relative importance of each component of the SPV to nodal
propagation influence. Typically, the first three components of the SPV
suffice to quantify the nodal influence. Generally, the nodal degree (the
first-order component of SPV) to some extent reflects the propagation
influence of nodes. However, the extent to which the two-, three- and
higher-dimensional structure of the SPV reflect the influence of nodes
is worth elucidating. For this purpose, we study the relative importance
of the two- and three-dimensional SPVs to nodal influence. In particular,
we fix $w_1=1$ and adjust $(w_2,w_3)$ in the range $0.1\leq (w_2,w_3)\leq 10$.
Figures~\ref{fig:Wei1} and \ref{fig:Wei2} show that, except for the CA-GrQc
network, when the SPV-based cluster centrality is used to quantify the nodal
influence, the optimal region is $w_2>1,w_3<1$. We also find that increasing
the weight of the second-order structural position has a larger impact on the
nodal influence than increasing the weight of the third-order structural
position. In the CA-GrQc network, the optimal area is in $w_3>1$, and
increasing the weight of the third-order structural position has a larger
impact on the ranking performance of cluster centrality. These results indicate
that, when the average path length of the network is small, the contribution
of the second-order adjacency relationship to the nodal influence is greater
than that of the third-order relationship. When the average path length of
the network is large (e.g., the CA-GrQc network whose average path length is
$\langle d \rangle=6.05$), the contribution of the node's third-order
adjacency relationship to the nodal influence is greater than that of the
second-order relationship. (The average path lengths of the six real-world
networks are listed in table~I in the main text).

\section*{Supplementary Note 4: General proof of main theorem: nodes with equal SPVs are symmetrical to each other}

Following the outline of the proof of the main theorem in {\bf Methods} in
the main text, here we provide a general and detailed proof. If the SPVs of
nodes $x$ and $y$ are equal, their corresponding eigencomponents satisfy the
following equations:
\begin{equation} \label{eq:SN4_1}
\left \{
\begin{array}{ll}
	\sum_{i=1}^N c_i \lambda_i(\eta_{i,x}-\eta_{i,y})=0,\\
	\sum_{i=1}^N c_i \lambda_i^2(\eta_{i,x}-\eta_{i,y})=0,\\
        \vdots\\
        \sum_{i=1}^N c_i \lambda_i^N(\eta_{i,x}-\eta_{i,y})=0,
\end{array}
\right.
\end{equation}
where some of the eigenvalues of the network adjacency matrix can be zero.
Without loss of generality, we arbitrarily assume $\lambda_1 =0$, $c_2 = 0$.
In general, a large network has repeated eigenvalues. If the $k$th and the
$(k+1)$th eigenvalues are equal: $\lambda_k=\lambda_{k+1}$,
Eq.~\eqref{eq:SN4_1} becomes
\begin{equation} \label{eq:SN4_2}
\left \{
\begin{array}{ll}
	\sum_{i=3}^{k-1} c_i \lambda_i(\eta_{i,x}-\eta_{i,y}) + \lambda_k(c_k(\eta_{k,x}-\eta_{k,y})+c_{k+1}(\eta_{k+1,x}-\eta_{k+1,y}) \\
	+ \sum_{i=k+2}^N c_i\lambda_i(\eta_{i,x}-\eta_{i,y})=0, \\
      \sum_{i=3}^{k-1} c_i \lambda_i^2(\eta_{i,x}-\eta_{i,y})+\lambda_k^2(c_k(\eta_{k,x}-\eta_{k,y})+c_{k+1}(\eta_{k+1,x}-\eta_{k+1,y}) \\
	+ \sum_{i=k+2}^N c_i\lambda_i^2(\eta_{i,x}-\eta_{i,y})=0,\\
      \vdots\\
      \sum_{i=3}^{k-1} c_i \lambda_i^{N}(\eta_{i,x}-\eta_{i,y})+\lambda_k^{N}(c_k(\eta_{k,x}-\eta_{k,y})+c_{k+1}(\eta_{k+1,x}-\eta_{k+1,y}) \\
	+\sum_{i=k+2}^{N} c_i\lambda_i^{N}(\eta_{i,x}-\eta_{i,y})=0.
\end{array}
\right.
\end{equation}
We take the first $(N-3)$ equations from Eq.~\eqref{eq:SN4_2} to form a new
set of equations. Regarding $\eta_{i,x}-\eta_{i,y}$ for
$i=3,4,\cdots,k-1,k+2,k+3,\cdots,N$ and
$c_k(\eta_{k,x}-\eta_{k,y})+c_{k+1}(\eta_{k+1,x}-\eta_{k+1,y})$
for $i=k,k+1$ as unknown quantities, we have that the determinant of the
coefficient of the new set of equations is
\begin{align}
C'&=\left|\begin{matrix}c_3\lambda_3 & c_4\lambda_4 & \cdots & c_{k-1} \lambda_{k-1} & \lambda_k & c_{k+2} \lambda_{k+2} & \cdots & c_N \lambda_N \\ c_3\lambda_3^2 & c_4\lambda_4^2 & \cdots & c_{k-1} \lambda_{k-1}^2 & \lambda_k^2 & c_{k+2} \lambda_{k+2}^2 & \cdots & c_N \lambda_N^2 \\ c_3\lambda_3^3 & c_4\lambda_4^3 & \cdots & c_{k-1} \lambda_{k-1}^3 & \lambda_k^3 & c_{k+2} \lambda_{k+2}^3 & \cdots & c_N \lambda_N^3  \\ \vdots & \vdots & \vdots & \vdots &\vdots & \vdots &\vdots & \vdots\\ c_3\lambda_3^{N-3} & c_4\lambda_4^{N-3} & \cdots & c_{k-1}\lambda_{k-1}^{N-3} & \lambda_k^{N-3} & c_{k+2}\lambda_{k+2}^{N-3} & \cdots & c_N\lambda_{N}^{N-3}\end{matrix} \right| \notag \\
&= c_3c_4\cdots c_{k-1}c_{k+2} \cdots c_N \lambda_3 \lambda_4\cdots \lambda_k \lambda_{k+2} \cdots \lambda_N \notag \\
	&\cdot\left|\begin{matrix}1 & 1 & \cdots & 1 & 1 & 1 & \cdots & 1 \\ \lambda_3 & \lambda_4 & \cdots &  \lambda_{k-1} & \lambda_k & \lambda_{k+2} & \cdots &  \lambda_N \\ \lambda_3^2 & \lambda_4^2 & \cdots &  \lambda_{k-1}^2 & \lambda_k^2 &  \lambda_{k+2}^2 & \cdots &  \lambda_N^2  \\ \vdots & \vdots & \vdots & \vdots &\vdots & \vdots &\vdots & \vdots\\ \lambda_3^{N-4} & \lambda_4^{N-4} & \cdots & \lambda_{k-1}^{N-4} & \lambda_k^{N-4} & \lambda_{k+2}^{N-4} & \cdots & \lambda_{N}^{N-4}\end{matrix} \right| \notag
\end{align}
Because the multiplicity of eigenvalues has been eliminated, we have
$\lambda_3 \neq \lambda_4 \neq  \cdots\neq \lambda_{k} \neq \lambda_{k+2} \neq \cdots \neq \lambda_N$ and $c_i\neq 0$ for $i=3,4,\cdots,k-1,k+2,k+3,\cdots,N$,
so $C'\neq 0$. The only solution of Eq.~\eqref{eq:SN4_2} is trivial. In the
eigenvectors representing $\mathbf{L}^0$, insofar as the corresponding
eigenvalues are not zero, the eigencomponents corresponding to nodes $x$ and
$y$ are equal.

For nodes $x$ and $y$ with equal SPVs, we further analyze the relationship
between their eigencomponents in the remaining eigenvectors. In particular,
by transforming and simplifying the characteristic equations
$\mathcal{A}\cdot\bm\eta=\lambda \bm\eta $, the eigencomponents corresponding
to nodes $x$ and $y$ can be represented by the eigencomponents of any identical
nodal set $S$. We assume that the eigencomponents corresponding to nodes $x$
and $y$ are represented by the eigencomponents of node $s_c$ and write
\begin{eqnarray}
\label{eq:SN4_3}
\varphi_0(\lambda_i) \eta_{i,x} & = & \varphi_1(\lambda_i) \eta_{i,s_c}, \\
\label{eq:SN4_4}
\psi_0(\lambda_i)\eta_{i,y} & = & \psi_1(\lambda_i) \eta_{i,s_c},
\end{eqnarray}
or
\begin{equation} \label{eq:SN4_5}
\frac{\varphi_0(\lambda_i)}{\psi_0(\lambda_i}) \eta_{i,x}
=\frac{\varphi_1(\lambda_i)}{\psi_1(\lambda_i)}\eta_{i,y}.
\end{equation}
Since $\varphi_0(\lambda_i)/\psi_0(\lambda_i)$ and
$\varphi_1(\lambda_i)/\psi_1(\lambda_i)$ are  polynomials in $\lambda_i$, we
let $T_0(\lambda_i)\equiv \varphi_0(\lambda_i)/\psi_0(\lambda_i)$ and
$T_1(\lambda_i) \equiv \varphi_1(\lambda_i)/\psi_1(\lambda_i)$.
Equation~\eqref{eq:SN4_5} becomes
\begin{equation} \label{eq:SN4_6}
T_0(\lambda_i)\eta_{i,x} = T_1(\lambda_i)\eta_{i,y}.
\end{equation}
If the eigenvalue $\lambda_i$ associated with the eigenvector used to
represent $\mathbf{L}^0$ is non-zero, we have $\eta_{i,x}=\eta_{i,y}$ and
\begin{equation} \label{eq:SN4_7}
T_0(\lambda_i) =T_1(\lambda_i).
\end{equation}
It is worth noting that the condition for Eq.~\eqref{eq:SN4_7} to hold does not
limit the value of $\lambda_i$ (except zero), indicating that the equality
$T_0(\lambda_i) =T_1(\lambda_i)$ does not depend on the value of $\lambda_i$.
That is, the polynomials $T_0(\lambda_i)$ and $T_1(\lambda_i)$ are exactly
the same. Substituting $\varphi_0(\lambda_i)= T_0(\lambda_i)\psi_0(\lambda_i)$
and $\varphi_1(\lambda_i)= T_1(\lambda_i)\psi_1(\lambda_i)$ into the
Eq.~\eqref{eq:SN4_3} gives
\begin{equation} \label{eq:SN4_8}
T_0(\lambda_i)\psi_0(\lambda_i) \eta_{i,x}=T_1(\lambda_i) \psi_1(\lambda_i) \eta_{i,s_c},
\end{equation}
or
\begin{equation} \label{eq:SN4_9}
\psi_0(\lambda_i) \eta_{i,x}=\psi_1(\lambda_i) \eta_{i,s_c}.
\end{equation}
From Eqs.~\eqref{eq:SN4_3} and \eqref{eq:SN4_9}, we see that the
eigencomponents corresponding to nodes $x$ and $y$ can be represented by the
eigencomponents of any identical nodal set $S$, and the form of representation
is exactly the same.

\begin{figure} [ht!]
\includegraphics[width=\linewidth]{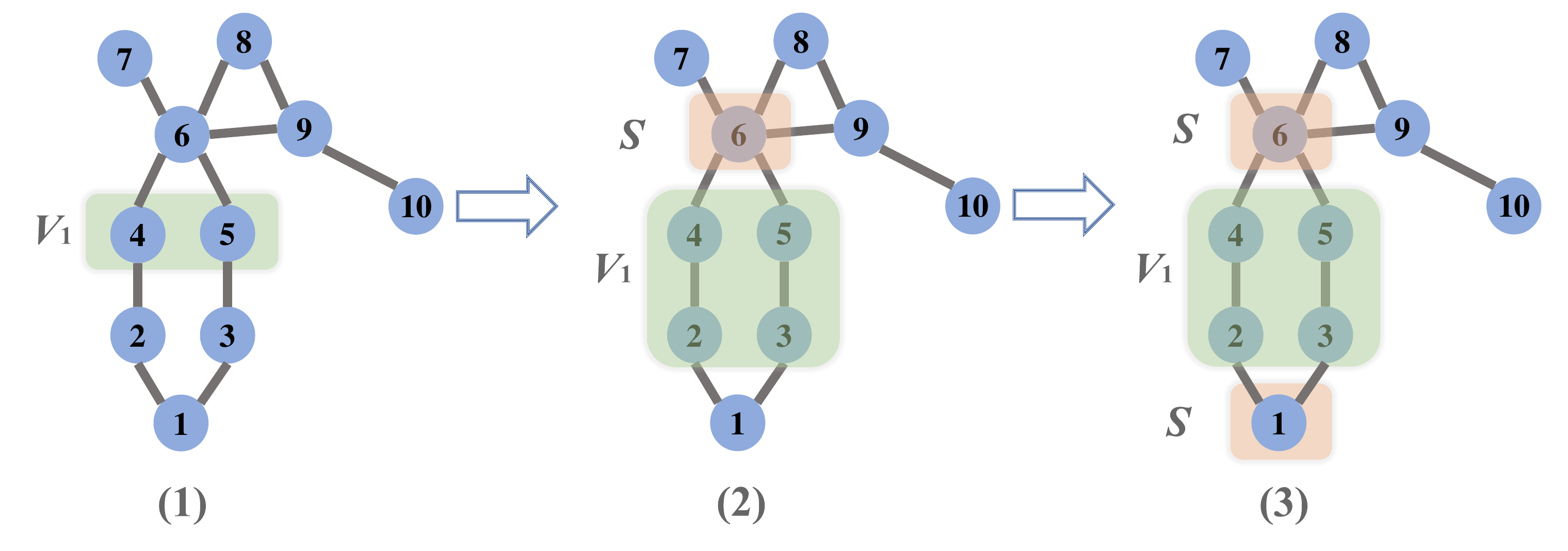}
\caption{
Schematic illustration of determining nodal sets $V_1$ and S.
(1) Take nodes 4 and 5 as an example. Let $V_1=\{4,5\}$. The characteristic
equations of the two nodes are
$\lambda_i \eta_{i,4}=\eta_{i,2}+\eta_{i,6}$ (e1) and
$\lambda_i \eta_{i,5}=\eta_{i,3}+\eta_{i,6}$ (e2), respectively.
(2) $S=\{6\}$. Substitute the characteristic equations of nodes 2 and 3 into
(e1) and (e2), respectively, to get
$(\lambda_i^2-1)\eta_{i,4}=\eta_{i,1}+\lambda_i \eta_{i,6}$,
$(\lambda_i^2-1)\eta_{i,5}=\eta_{i,1}+\lambda_i \eta_{i,6}$, and
$V_1=\{2,3,4,5\}$. (3) $S = \{1,6\}$. The characteristic components of nodes
4 and 5 can be represented by the characteristic components of node set $S$,
where $V_1=\{2,3,4,5\}$ and $S = \{1,6\}$.}
\label{fig:Sch}
\end{figure}

To find the identical node set $S$, we define the characteristic equation of
node $j$ as $\lambda_i\eta_{i,j} = \sum_{l \in \Lambda_j}\eta_{i,l}$, where
$i=1,2,\cdots,N$ and $\Lambda_j$ is the set of neighboring nodes of $j$. That
is, the eigencomponent of node $j$ can be expressed by the eigencomponents of
all its neighboring nodes. The process, as illustrated in Fig.~\ref{fig:Sch},
consists of the following four steps.
\begin{enumerate}
\item
Analyze the characteristic equation
$\lambda_i \eta_{i,x} =\sum_{j=1}^{k_x} \eta_{i,x_j}$ of node $x$ with degree
$k_x$, where $x_j$ is a neighboring node of $x$. If all neighbors of $x$ are
removed, node $x$ will be separated from the network. Denote $V_1=\{x\}$.
\item
Substitute the characteristic equation of a neighboring node $x_1$ of $x$
into the characteristic equation of $x$ so that the eigencomponents of node
$x$ are represented by the eigencomponents of its neighbors (excluding $x_1$)
and the neighbors of $x_1$. If the nodes used to represent the eigencomponents
of $x$ are removed, the local structure formed by nodes $x$ and $x_1$ must be
separated from the network. Denote $V_1=\{x,x_1\}$.
\item
Continue the iterative process in step 2 until the eigencomponents of node
$x$ is represented by the eigencomponents of nodal set $S$. If the
characteristic equation of node $m$ is used for iteration, then $m$ is added
to $V_1$ and $V_1=\{x,x_i,\cdots\}$. Remove nodal set $S$. The local
structure $\mbox{local}_G(V_1,E_1)$ formed by $V_1$ is separated from the
network, where $E_1$ is the set of internal edges of the node set $V_1$.
\item
Two possible situations can arise in determining nodal set $S$: two
nodes may or may not have common neighboring nodes of different orders.
The first case is where nodes $x$ and $y$ have common neighboring nodes. In
this case, simplify the characteristic equations of nodes $x$ and $y$
according to step 2. If $x$ and $y$ have a common neighboring set $s_{1,xy}$,
then $S=\{s_{1,xy}\}$. In this case, simplify the characteristic equations
of $x$ and $y$ with the characteristic equations of non-common neighboring
nodes and add these nodes to the set $V_1$. If there is a second-order common
neighboring set $s_{2,xy}$ between nodes $x$ and $y$, add them to the set $S$
to get $S=\{s_{1, xy},s_{2,xy}\}$. Continue the process until the
eigencomponents of $x$ and $y$ can be represented by the eigencomponents of
$S$. The second case is where nodes $x$ and $y$ have no $n$th-order common
neighbors. In this case, all neighbors of $x$ constitute nodal set $S$.
\end{enumerate}

Let the identical nodal set be $S=\{s_1,s_2,\cdots,s_k\}$. Nodes of the network
can then be divided into three sets: $V_1$, $S$, and the remaining nodal set
$V\verb|/|\{V_1,S\}$. We perform elementary row transformation on matrix
$\mathcal{A}$ and divide it into blocks by row to get
\begin{equation} \label{eq:SN4_12}
\mathcal{A}\cdot\bm\eta =
\left[\begin{matrix}
\mathcal{A}_1\\
\mathcal{A}_2\\
\mathcal{A}_3
\end{matrix}\right]
\cdot\bm\eta = \lambda\bm\eta,
\end{equation}
where $\mathcal{A}_1$, $\mathcal{A}_2$ and $\mathcal{A}_3$ are, respectively,
${\vert V_1\vert \times N}$, ${\vert S\vert \times N}$, and
${\vert V\verb|/|\{V_1,S\} \vert \times N}$ adjacency matrices of the local
structures formed by nodal sets $V_1$, $S$, $V\verb|/|\{V_1,S\}$ together
with their neighbors. This analysis indicates that the eigencomponents of
nodes $x$ and $y$ can be represented by the eigencomponents of nodal set
$S=\{s_1,s_2,\cdots,s_k\}$ through the characteristic equations
$\mathcal{A}_1\bm\eta=\lambda \bm\eta$. Since we have proved that the
eigencomponents corresponding to $x$ and $y$ can be represented by the
eigencomponents of any identical nodal set $S$, the representation is exactly
the same. We thus have
\begin{eqnarray}
\label{eq:SN4_13}
	f_0(\lambda_i)\eta_{i,x} & = & f_1(\lambda_i)\eta_{i,s_1}+f_2(\lambda_i)\eta_{i,s_2}+\cdots+f_k(\lambda_i)\eta_{i,s_k}, \\
\label{eq:SN4_14}
	f_0(\lambda_i)\eta_{i,y} & = & f_1(\lambda_i)\eta_{i,s_1}+f_2(\lambda_i)\eta_{i,s_2}+\cdots+f_k(\lambda_i)\eta_{i,s_k}.
\end{eqnarray}
If $f_0(\lambda_i) \neq0$, then $\eta_{i,x}=\eta_{i,y}$. Permuting nodes $x$ and
$y$, we have that the eigenvectors associated with such eigenvalues satisfy
$P\bm\eta=\bm\eta$.

We now discuss the case $f_0(\lambda_i) = 0$. It can be seen from
Eqs.~\eqref{eq:SN4_13} and \eqref{eq:SN4_14} that the eigencomponents of nodes
$x$ and $y$ are free variables. In order to find the relationship between
the eigencomponents of $x$ and $y$, we further analyze the characteristic
equation $\mathcal{A}\cdot\bm\eta=\lambda\bm\eta$. To do this, we move $x$
and $y$ out of nodal set $V_1$ and divide all nodes in the network into
two sets: $V_1 \verb|/| \{x,y\}$ and $\{V\verb|/|V_1,x,y\}$. The characteristic
equation can be written as
\begin{equation} \label{eq:SN4_15}
	\mathcal{A}\cdot\bm\eta=\left[\begin{matrix} \mathcal{A}_1' \\ \mathcal{A}_2' \end{matrix}\right]\cdot\bm\eta=\lambda \bm\eta ,
\end{equation}
where $\mathcal{A}_1'$ and $\mathcal{A}_2'$ are, respectively,
$\lvert V_1 \verb|/| \{x,y\} \rvert \times N$- and
$\vert \{V \verb|/|V_1,x,y\}\vert \times N$-dimensional
adjacency matrices of the local structures formed by nodal sets
$V_1 \verb|/| \{x,y\}$ and $\{V \verb|/|V_1,x,y\}$ as well as their neighbors.
The eigencomponents of $V_1 \verb|/|\{x,y\}$ can be represented by the
eigencomponents of the set $\{x,y,S\}$ through the characteristic equation
$\mathcal{A}_1'\bm\eta = \lambda \bm\eta$. From Eqs.~\eqref{eq:SN4_13} and
\eqref{eq:SN4_14}, the eigencomponents of $x$ and $y$ are related to the
eigencomponents of nodal set $S$.

We are thus led to analyze the characteristic equation of any node $s_l\in S$:
$\lambda_i \eta_{i,s_l} =\sum_{j\in \Lambda_{s_l}}\eta_{i,j}$, where
$i =1,2,\cdots,N$ and $\Lambda_{s_l}$ is the set of neighbors of node $s_l$.
We divide the neighbors of $s_l$ into three parts: the first part is a subset
of nodal set $V_1$, the second part belongs to the set $S$, and the
third part is contained in the set $V \verb|/|\{V_1,S\} $. Since the
eigencomponents of the nodes belonging to the set $V_1$ can be represented
by the eigencomponents of the set $\{x,y,S\}$, the characteristic equation
of node $s_l$ can be written as
\begin{equation} \label{eq:SN4_16}
F_0(\lambda_i)\eta_{i,s_l}=G_0(\lambda_i)\eta_{i,x}+G_1(\lambda_i)\eta_{i,y}+\sum_{s_j \in S \verb|/| s_l} F_j(\lambda_i)\eta_{i,s_j}+\sum_{j=1}^m H_j(\lambda_i)\eta_{i,r_j},
\end{equation}
where nodal set $R(r_1,r_2,\cdots,r_m)$ are the neighbors of node $s_l$
in the set $V \verb|/|\{V_1,S\}$. Rearranging the terms, we get
\begin{equation} \label{eq:SN4_17}
F_0(\lambda_i)\eta_{i,s_l}-\sum_{s_j \in S \verb|/| s_l} F_j(\lambda_i)\eta_{i,s_j}-\sum_{j=1}^m H_j(\lambda_i)\eta_{i,r_j}=G_0(\lambda_i)\eta_{i,x}+G_1(\lambda_i)\eta_{i,y}.
\end{equation}
Let $E\equiv F_0(\lambda_i)\eta_{i,s_l}-\sum_{s_j \in S \verb|/| s_l} F_j(\lambda_i)\eta_{i,s_j}-\sum_{j=1}^m H_j(\lambda_i)\eta_{i,r_j}$.
Equation~\eqref{eq:SN4_17} leads to
\begin{eqnarray}
\label{eq:SN4_18}
	G_0(\lambda_i)\eta_{i,x} & = & E-G_1(\lambda_i)\eta_{i,y}, \\
\label{eq:SN4_19}
	G_1(\lambda_i)\eta_{i,y} & = & E-G_0(\lambda_i)\eta_{i,x}.
\end{eqnarray}
Multiplying both sides of Eqs.~\eqref{eq:SN4_13} and \eqref{eq:SN4_14}
by $G_0(\lambda_i)$ and $G_1(\lambda_i)$, respectively, and substituting
Eqs.~\eqref{eq:SN4_18} and \eqref{eq:SN4_19} into the resulting equations,
we get
\begin{eqnarray}
\label{eq:SN4_20}
	f_0(\lambda_i)G_1(\lambda_i)\eta_{i,y} & = & f_0(\lambda_i)E-(f_1(\lambda_i)\eta_{i,s_1}+f_2(\lambda_i)\eta_{i,s_2}+\cdots+f_k(\lambda_i)\eta_{i,s_k})G_0(\lambda_i), \\
\label{eq:SN4_21}
	f_0(\lambda_i)G_0(\lambda_i)\eta_{i,x} & = & f_0(\lambda_i)E-(f_1(\lambda_i)\eta_{i,s_1}+f_2(\lambda_i)\eta_{i,s_2}+\cdots+f_k(\lambda_i)\eta_{i,s_k})G_1(\lambda_i).
\end{eqnarray}
Since we have proved that the eigencomponents corresponding to nodes $x$ and
$y$ can be represented by the eigencomponents of any identical nodal set, and
the way of representation is exactly the same, we have
$G_0(\lambda_i)=G_1(\lambda_i)$. Equation~\eqref{eq:SN4_16} can be further
simplified to
\begin{equation} \label{eq:SN4_22}
F_0(\lambda_i)\eta_{i,s_l}=G_0(\lambda_i)(\eta_{i,x}+\eta_{i,y})+\sum_{s_j \in S \verb|/| s_l} F_j(\lambda_i)\eta_{i,s_j}+\sum_{j=1}^m H_j(\lambda_i)\eta_{i,r_j}.
\end{equation}
The above analysis indicates that Eqs.~\eqref{eq:SN4_13} and \eqref{eq:SN4_14}
are actually the consequence of the characteristic equations
$\mathcal{A}_1\bm\eta=\lambda\bm\eta$. As a result, the eigenvalue satisfying
$f_0(\lambda_i)=0$ is determined by the local structure $\mbox{local}_G(V_1,E_1)$
of nodal set $V_1$ only. Likewise, the eigenvector associated with this
eigenvalue is related to the local structure $\mbox{local}_G(V_1,E_1)$ formed
by $V_1$ only, so the eigencomponents of nodal set $V \verb|/|V_1$ can be
directly set to zero. Substituting the values of these components into the
Eq.~\eqref{eq:SN4_22}, we have that the left side of this equation is zero, so
are the second and third terms on the right side. We thus have
\begin{displaymath}
	G_0(\lambda_i)(\eta_{i,x}+\eta_{i ,y})=0 \ \ \mbox{or} \ \
	\eta_{i,x} +\eta_{i,y}=0.
\end{displaymath}
Apparently, for $f_0(\lambda_i) \neq 0$, the local structure
$\mbox{local}_G(V_1,E_1)$ is not sufficient to determine this eigenvalue, so
it must be related to the structure outside nodal set $V_1$. The
eigenvectors of matrix $\mathcal{A}$ can be classified into two categories.
The first category is $f_0(\lambda_i) \neq 0$, the eigenvectors are
determined by the eigencomponents not only of nodal set $V_1$ but also
of the set $V\verb|/|V_1$. The eigencomponents of nodes $x$ and $y$ satisfy
$\eta_{i,x}=\eta_{i,y}$. In the second category defined by $f_0(\lambda_i) = 0$,
the eigenvectors are related to the eigencomponents of nodal set $V_1$
only, and the eigencomponents of nodal set $V\verb|/|V_1$ are all zero.
In this case, the eigencomponents of $x$ and $y$ satisfy
$\eta_{i,x}+\eta_{i,y}=0$.

If there are nodes $x'$ and $y'$ with the same SPV in nodal set
$V_1\verb|/|\{x,y\}$, their characteristic components have similar properties
to those of nodes $x$ and $y$. In particular, for the first type of
eigenvectors, the eigencomponents of $x'$ and $y'$ satisfy
$\eta_{i,x'}=\eta_{i,y'}$. For the second type, we have
$\eta_{i,x'}+\eta_{i,y'}=0$. If there is a node $x''$ in nodal set
$V_1\verb|/|\{x,y\}$, whose SPV is not equal to that of any other node
in the set, then by using the characteristic equations
$\mathcal{A}_1\bm\eta=\lambda \bm\eta$, the eigencomponents of $x''$ are
represented by those of nodal set $S$. Similarly, consider the
eigencomponents of node $x''$ for the two types of eigenvectors. In the
first type, the value of the eigencomponent of $x''$ can be arbitrary.
In the second type, according to Eq.~\eqref{eq:SN4_16} or
Eq.~\eqref{eq:SN4_22}, we have $\eta_{i,x''}=0$. If the nodes $\{x,x'\}$
are permuted with $\{y,y'\}$, respectively, which have the same SPVs,
all the eigenvectors of matrix $\mathcal{A}$ satisfy $P\bm\eta=\bm\eta$
or $P \bm\eta=-\bm\eta$. If nodal set $V_1$ contains nodes $x$ and $y$
with the same SPV only, we permute the two nodes so that all the eigenvectors
of matrix $\mathcal{A}$ still satisfy $P\bm\eta=\bm\eta$ or
$P\bm\eta=-\bm\eta$. In the following, we assume that nodal set $V_1$
contains only the nodes with equal SPVs.

When there are $K$ ($K>2$) nodes with the same SPV, we have
$V_1=\{q_1,q_2,\cdots,q_K\}$. For the first type of eigenvectors, the
eigencomponents of these $K$ nodes satisfy
$\eta_{i,q_1}=\eta_{i,q_2}=\cdots=\eta_{i,q_K}$. Arbitrarily permuting these
$K$ nodes implies that the eigenvectors satisfy $P\bm\eta=\bm\eta$.
For the second type of eigenvectors, the eigencomponents of the $K$ nodes
satisfy $\sum_{j=1}^K \eta_{i,q_j}=0$. For $K=2n$, where $n$ is a positive
integer greater than one, this gives $2n-1$ linearly independent eigenvectors.
From nodal set $\{q_1,q_2,\cdots,q_K\}$, we pair the nodes arbitrarily.
Without loss of generality, we write the pairing sequence as
$(q_1,q_2),(q_3,q_4),\dots,(q_{2j-1},q_{2j}),\cdots,(q_{2n-1},q_{2n})$. We
then permute these paired nodes. We can prove that, with permutation under
the above rule, all the eigenvectors of matrix $\mathcal{A}$ satisfy
$P\bm\eta=\pm \bm\eta$. In particular, the first type of eigenvectors satisfy
$P\bm\eta=\pm\bm\eta$, so it is only necessary to prove that permuting the
nodes can lead to $P\bm\eta=\pm\bm \eta$ for the second type of eigenvectors.
We thus have that the number of eigenvectors of the second type is $2n-1$.
We then need to construct $2n-1$ eigenvectors that satisfy
$P\bm\eta=\pm\bm \eta$.

For the $j$th eigenvector ($j=1,2,\cdots,n$), let
$(\eta_{j,q_{2j-1}},\eta_{j,q_{2j}})=(1,-1)$ and the eigencomponents of the
remaining $K-2$ nodes in nodal set $V_1$ be zero. We obtain $n$
$K$-dimensional vectors. Since, for the second type of eigenvectors, except
for nodal set $V_1$, the eigencomponents of the rest of the nodes are zero,
we can simply expand them into $|V|$-dimensional eigenvectors as
\begin{eqnarray}
\nonumber
	&& \left( \begin{matrix} 1 & -1 & 0 & 0 & \cdots & 0 & 0 & 0 & 0 \\0 & 0    &1   & -1 &\cdots & 0 & 0 & 0& 0\\ \vdots & \vdots & \vdots & \vdots & \vdots & \vdots & \vdots & \vdots & \vdots \\ 0 & 0 & 0 & 0 & \cdots & 1 & -1 & 0 & 0\\ 0 & 0 & 0 & 0 & \cdots & 0 & 0 & 1 & -1 \end{matrix} \right) \Longrightarrow \\
\label{eq:SN4_23}
	&& \left(\begin{matrix} 1 & -1 & 0 & 0 & \cdots & 0 & 0 & 0 & 0 \\0 & 0    &1   & -1 &\cdots & 0 & 0 & 0& 0\\ \vdots & \vdots & \vdots & \vdots & \vdots & \vdots & \vdots & \vdots & \vdots \\ 0 & 0 & 0 & 0 & \cdots & 1 & -1 & 0 & 0\\ 0 & 0 & 0 & 0 & \cdots & 0 & 0 & 1 & -1 \\ \end{matrix} \left.\rule{0mm}{15mm} \right | \begin{matrix} 0& 0& \cdots &0 &0\\ 0 & 0 & \cdots &0 & 0\\ \vdots & \vdots & \ddots & \vdots & \vdots \\0 & 0 &\cdots &0 & 0 \\ 0 & 0 & \cdots & 0 & 0\end{matrix} \right).
\end{eqnarray}
The $n$ eigenvectors obtained by this method are linearly independent, which
can be proved, as follows.

\underline{Proof}: Denote each row of the matrix on the right side of
\eqref{eq:SN4_23}, from top to bottom, as $\eta_1,\eta_2,\cdots,\eta_n$,
respectively. There exist $k_1,k_2,\cdots,k_n$ such that
$k_1\eta_1+k_2\eta_2+\cdots+ k_n\eta_n=0$. We then have
\begin{equation} \label{eq:SN4_24}
(k_1,-k_1,k_2,-k_2,\cdots,k_n,-k_n,0,0,\cdots,0,0)=(0,0,0,0,\cdots,0,0,0,0,\cdots,0,0).
\end{equation}
We have $k_1 = k_2=\cdots=k_n=0$, indicating that the group of vectors
$\eta_1,\eta_2,\cdots,\eta_n$ are linearly independent, which are $n$
linearly independent eigenvectors that satisfy $P\bm\eta=-\bm\eta$.
$\blacksquare$

We search for $n-1$ linearly independent eigenvectors that satisfy
$P\bm\eta=\bm\eta$, which are also linearly independent of the $n$
eigenvectors found above. For the $j$th eigenvector ($j=1,2,\cdots,n$), let
$(\eta_{j,q_1},\eta_{j,q_2})=(1,1),(\eta_{j,q_{2j-1}},\eta_{j,q_{2j}})=(-1,-1)$
and the eigencomponents of the remaining $K-4$ nodes in nodal set $V_1$ be
zero. This leads to $n-1$ $K$-dimensional vectors. Similarly, we expand them
into $|V|$-dimensional eigenvectors, as in formula \eqref{eq:SN4_23}.
Apparently, the eigenvectors so constructed satisfy $P\bm\eta=\bm\eta$.
Here we prove that the newly constructed $n-1$ eigenvectors are linearly
independent of each other and are also linearly independent of the $n$
eigenvectors found
above.

\underline{Proof}: The matrix composed of the newly constructed $n-1$ vectors is
\begin{equation} \label{eq:SN4_25}
\left(\begin{matrix} 1 & 1 & -1 & -1 & 0 & 0 & \cdots & 0 & 0 & 0 & 0 \\1 & 1    & 0   & 0 & -1 & -1 & \cdots & 0 & 0 & 0& 0\\ \vdots & \vdots & \vdots & \vdots & \vdots & \vdots & \ddots & \vdots & \vdots & \vdots & \vdots \\ 1 & 1 & 0 & 0 & 0 & 0 & \cdots & -1 & -1 & 0 & 0 \\ 1 & 1 & 0 & 0 & 0 & 0 & \cdots & 0 & 0 & -1 & -1 \\ \end{matrix} \left.\rule{0mm}{15mm}\right |   \begin{matrix} 0 & 0 & \cdots & 0 & 0\\ 0 & 0 & \cdots & 0 & 0\\ \vdots & \vdots & \ddots & \vdots & \vdots \\ 0 & 0 &\cdots &0 & 0 \\ 0 & 0 & \cdots & 0 & 0\end{matrix}  \right).
\end{equation}
Denote each row of the matrix as $\eta_{n+1},\eta_{n+2},\cdots,\eta_{2n-1}$,
respectively, from top to bottom. There exist $k_1,k_2,\cdots,k_{2n-1}$ such
that $k_1\eta_1+k_2\eta_2+\cdots+ k_{2n-1}\eta_{2n-1}=0$. We have
\begin{equation} \label{eq:SN4_26}
\left \{
\begin{array}{rl}
     k_1+\sum_{l=n+1}^{2n-1}k_l=0,    \\
     -k_1+\sum_{l=n+1}^{2n-1}k_l=0,
\end{array}
\right.
\end{equation}
and
\begin{equation} \label{eq:SN4_27}
\left \{\begin{aligned} k_{j+1}-k_{n+j}=0    \\ -k_{j+1}-k_{n+j}=0 \end{aligned} \right.,j=1,2,\cdots,n-1.
\end{equation}
From Eqs.~\eqref{eq:SN4_26} and \eqref{eq:SN4_27}, we have
$k_1 = k_2=\cdots=k_{2n-1}=0$, which means that the group of vectors
$\eta_1,\eta_2,\cdots,\eta_{2n-1}$ are linearly independent of each other,
leading to $2n-1$ linearly independent eigenvectors. Among them, $n$
eigenvectors satisfy $P\bm\eta=-\bm\eta$ and $n-1$ eigenvectors satisfy
$P\bm\eta=\bm\eta$. $\blacksquare$

For $K=2n+1$, where $n$ is a positive integer, there are $2n$ linearly
independent eigenvectors. Similarly, we perform node pairing and permutation
to construct the $2n$ eigenvectors. The first $2n-1$ eigenvectors are
constructed in the same way as for $K=2n$, and the last eigenvector is
$\eta_{2n}=(1,1,\cdots, 1,1,-2n,0,0\cdots,0,0)$. The eigenvector $\eta_{2n}$
is then linearly independent of the above $2n-1$ eigenvectors and we have
$P\bm\eta=\bm\eta$.

To summarize this Supplementary Note, we have that, according to the
permutation method explained, all eigenvectors of the network satisfy
$P\bm\eta=\pm \bm\eta$. Since we pair nodes arbitrarily, permuting any two
nodes can make all the eigenvalues satisfy $P\bm\eta=\pm \bm\eta$. According
to Lemma 2 in the main text, the permutation corresponding to the permutation
matrix $P$ is an automorphism. That is, nodes with identical SPVs are
symmetric to each other.

\bibliographystyle{naturemag}
\bibliography{Quantifyingsymmetry}